\documentstyle{mn}

\def\simless{\mathbin{\lower 2pt\hbox
   {$\rlap{\raise 5pt\hbox{$\char'074$}}\mathchar"7218$}}}   
\def\simgreat{\mathbin{\lower 2pt\hbox
   {$\rlap{\raise 5pt\hbox{$\char'076$}}\mathchar"7218$}}}   
\def\be {\begin{equation}}
\def\ee {\end{equation}}

\input psfig

\pubyear{2000}

\title[The properties of binary systems]{Predicting the properties of binary stellar systems: \\ the evolution of accreting protobinary systems}

\author[M. R. Bate]
  {Matthew R. Bate$^{1,2}$\thanks{E-mail: mbate@ast.cam.ac.uk} \\
  $^{1}$ Institute of Astronomy, Madingley Road, Cambridge CB3 0HA \\
  $^{2}$ Max-Planck-Institut f\"ur Astronomie, K\"onig\-stuhl 17, D-69117 Heidelberg, Germany}

\date{Accepted 1999 December 23. Received in original form 1999 September 16}  

\begin{document}
\label{firstpage}
\maketitle

\begin{abstract}

We investigate the formation of binary stellar systems.
We consider a model where a `seed' protobinary system forms, via
fragmentation, within a collapsing molecular cloud core and evolves
to its final mass by accreting material from an infalling 
gaseous envelope.  This accretion alters the mass ratio and orbit
of the binary, 
and is largely responsible for forming the circumstellar and/or 
circumbinary discs.

Given this model for binary formation, we predict the properties of 
binary systems and how they depend on the initial conditions within 
the molecular cloud core.  We predict that there should
be a continuous trend such that closer binaries are more likely 
to have equal mass components and are more likely to have 
circumbinary discs than wider systems.  Comparing our results to observations, 
we find that the observed mass-ratio distributions of binaries and the
frequency of circumbinary discs as a function of separation are most easily
reproduced if the progenitor molecular cloud cores have radial density
profiles between uniform and $1/r$ (e.g. Gaussian) with near 
uniform-rotation.  This is in good agreement with the observed properties
of pre-stellar cores.  Conversely, we find that the observed 
properties of binaries cannot be 
reproduced if the cloud cores are in solid-body rotation and 
have initial density profiles which are strongly centrally condensed. 
Finally, in agreement with the radial-velocity 
searches for extra-solar planets, we find that it is very 
difficult to form a brown dwarf companion 
to a solar-type star with a separation $\simless 10$ AU, but that the
frequency of brown dwarf companions should increase with larger
separations or lower mass primaries.

\end{abstract}


\section{Introduction}
\label{introduction}

The favoured mechanism for producing most binary stellar systems
is the fragmentation of a molecular cloud core during its gravitational
collapse.  Fragmentation can be divided into two main classes: 
direct fragmentation (e.g.~Boss \& Bodenheimer 1979; Boss 1986; 
Bonnell et al.~1991, 1992; Bonnell \& Bastien 1992;  
Nelson \& Papaloizou 1993; Burkert \& Bodenheimer 1993; Bate \& Burkert 1997), 
and rotational fragmentation 
(e.g.~Norman \& Wilson 1978; Bonnell 1994; Bonnell \& Bate 1994a, 1994b;
Burkert \& Bodenheimer 1996; Burkert, Bate, Bodenheimer 1997).  
Direct fragmentation depends critically on 
the initial density structure within the molecular cloud core 
(e.g.~non-spherical shape or density perturbations), whereas rotational
fragmentation is relatively independent of the initial 
density structure of the cloud because the fragmentation occurs 
due to nonaxisymmetric instabilities in a massive 
rotationally-supported disc or ring.

The main conclusion, from $\approx 20$ years of fragmentation studies,
is that it appears to be possible to form binaries with similar 
properties to those that are observed.  
However, it has not been possible to use these calculations to 
predict the fundamental properties of stellar systems such as
the fraction of stellar systems which are binary or the properties of binary 
systems (e.g.~the distributions of mass ratios,
separations, and eccentricities and the properties of 
discs in pre-main-sequence systems).  

There are two primary reasons for this lack of predictive power.  
First, the results of fragmentation calculations depend sensitively
on the initial conditions, which are poorly constrained.
The second problem is that of accretion.  In fragmentation calculations,
the binary or multiple protostellar systems that
are formed initially contain only a small fraction of the total mass of the
original cloud (e.g.~Boss 1986; Bonnell \& Bate 1994b)
with the magnitude of this fraction decreasing with the 
binary's initial separation (see Section \ref{bmass_vs_sepsec}).  
To obtain the final parameters of a stellar system, a calculation must
be followed until all of the original cloud material has been
accumulated by one of the protostars or their discs.
Unfortunately, due to the enormous range in densities and dynamical
time-scales in such a calculation, this is very difficult.
Thus far, only one calculation has followed the three-dimensional
collapse of a molecular cloud core until $>90$\% of the initial
cloud was contained in the protostars or circumstellar/circumbinary discs
(Bate, Bonnell \& Price 1995).
Because such calculations are so difficult to perform, it is
impossible to perform the number of calculations that would be
required to predict the statistical properties of 
binary stellar systems -- even if we knew the distribution of  
initial conditions.  On the other hand, if we can overcome this second
difficulty, we can use observations of binary systems to better constrain
the initial conditions for star formation.

Bate \& Bonnell \shortcite{BatBon97} quantified how 
the properties of a binary system are affected by the accretion 
of a small amount of gas from an infalling gaseous envelope.  
They found that the effects depend primarily on the specific 
angular momentum of the gas and the binary's mass ratio 
(see also Artymowicz 1983; Bate 1997).  Generally, accretion of gas 
with low specific angular momentum decreases the mass ratio and 
separation of the binary, while accretion of gas with high 
specific angular momentum increases the separation, drives the 
mass ratio toward unity, and can form a circumbinary disc.  
From these results, they predicted that closer binaries 
should have mass ratios that are biased toward equal masses compared
to wider systems.

In this paper, we use the results of Bate \& Bonnell \shortcite{BatBon97}
to develop a protobinary evolution code that enables us to 
follow the evolution of a protobinary system as it accretes 
from its initial to its final mass, but does so in far less time 
than would be required for a full hydrodynamic calculation.  
Using this code, we consider the following model for the formation of
binary stellar systems.  We assume that a `seed' binary system is 
formed at the centre of a collapsing molecular cloud core, presumably 
via some sort of fragmentation.  The protobinary system initial consists 
of only a small fraction of the total mass of the core.  
Subsequently, it accretes the remainder of the initial cloud (which 
is falling on to the binary) and its properties evolve due to the 
accretion.  We consider the formation process to be complete 
when all of the original cloud's material is contained
either in one of the two stars or their surrounding discs.
Our goal is to obtain
predictions about the properties of binary stars that can be tested
observationally, and to determine how these properties depend on the 
initial conditions (e.g.~the density and angular momentum profiles) 
in the progenitor molecular cloud cores
so that the initial conditions can be better constrained.

In Section \ref{commethod}, we describe the methods used to follow the
evolution of accreting protobinary systems, and we present the
results of various test calculations in Section \ref{comparison}.
In Sections \ref{evolution} and \ref{relax}, we present results from
calculations with a range of initial conditions which follow the 
evolution of accreting protobinary systems.
From these results, in Section \ref{predictions}, 
we make predictions regarding the properties of binary systems and compare
them with the latest observations.
These predictions are briefly summarised in Section \ref{conclusions}.

Those readers more interested in our predictions of the
properties of binary stars, rather than the method by which these 
predictions have been obtained, may care to move directly 
to Section \ref{evolution}.

\section{Computational methods}
\label{commethod}

\subsection{Protobinary evolution code}
\label{PBEcode}

Bate \& Bonnell \shortcite{BatBon97} considered the effects that 
accretion of a small amount of gas from an infalling gaseous envelope has
on the properties of a protobinary system.  They quantified
these effects as functions of the mass ratio of the protobinary
and the specific angular momentum of the infalling gas.
Therefore, if a `seed' binary system is formed at the centre
of a collapsing gas cloud, and we know its initial mass ratio and 
separation, the mass infall rate on to the 
binary, and the distribution of the specific angular momentum of the gas,
then we can determine how the binary will evolve as it accretes 
infalling gas.  Essentially, we can integrate the binary from its 
initial mass to its final mass by accreting the gas in a series 
of small steps and altering the masses of the binary's components and 
their orbit by the amounts determined by Bate \& Bonnell 
\shortcite{BatBon97} after each step.

\begin{figure}
\centerline{\hspace{0.0truecm}\psfig{figure=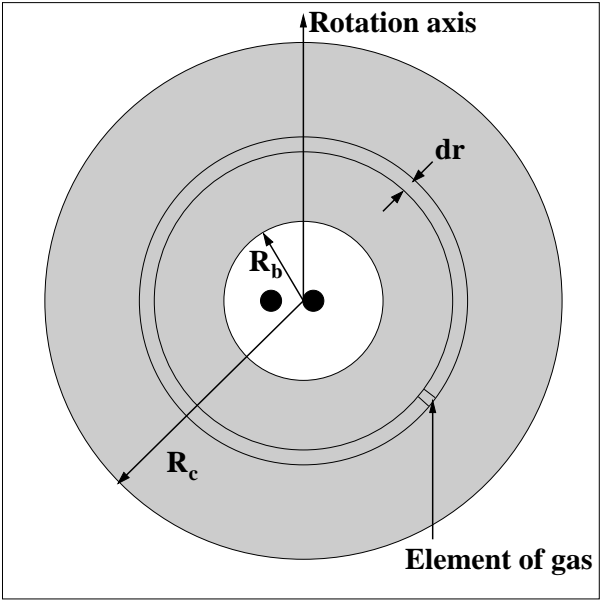,width=8.0truecm,height=8.0truecm,rwidth=8.0truecm,rheight=8.5truecm}}
\caption{\label{model}  The model for binary star formation which is considered in this paper (see Section 2.1).}
\end{figure}

We now describe the implementation of the protobinary evolution (PBE) code.
We begin with the properties of the molecular cloud core, before it 
begins to collapse dynamically, from which the binary system will form.
For simplicity, the progenitor cloud (see Figure \ref{model}) is assumed to be 
spherical with mass $M_{\rm c}$, radius $R_{\rm c}$, density distribution
\be
\rho = \rho_{\rm 0} \left(r/R_{\rm c}\right)^{\lambda},
\ee
and angular velocity
\be
\Omega_{\rm c} = \Omega_{\rm 0} \left(r/R_{\rm c}\right)^{\beta},
\ee
where $\lambda$ controls the initial central condensation of the cloud,
and $\beta$ gives the amount of differential rotation of the cloud.  Note
that $\Omega_{\rm c}$ is constant on spheres ($r$), 
not cylinders ($r_{\rm xy}$).  
Note also that if molecular cloud cores spend a large fraction of 
their lifetimes as magnetically-supported, quasistatic structures 
before undergoing dynamic collapse, they are likely to be in 
solid-body rotation and significantly centrally-condensed 
before the dynamic collapse begins.

We assume that a `seed' binary is formed at the centre of this 
molecular cloud core after it begins to collapse.
The `seed' binary contains only a small fraction of the total mass 
of the cloud.  The masses of the primary and secondary are 
$M_1$ and $M_2$, respectively.   The binary has a total mass $M_{\rm b}$, 
mass ratio $q=M_2/M_1$, separation $a$ and is in a circular orbit.  
We assume the binary has a circular orbit and that its axis of 
rotation is aligned with that of the cloud, since the PBE code makes 
use of the results of Bate \& Bonnell \shortcite{BatBon97} and they 
only studied such systems.

The `seed' binary is assumed to have formed from the gas that was 
in a spherical region at the centre of the progenitor 
cloud of radius $R_{\rm b}$ (Figure \ref{model}).  
Unless otherwise stated, we assume that this spherical region initially
had the same mass and angular momentum as the `seed' binary.
Thus, the initial angular momentum of the binary is given by
\be
\label{lbinary}
L_{\rm b} = \sqrt{G M_{\rm b}^3 a}~ \frac{q}{(1+q)^2} = \Lambda M_{\rm b} R_{\rm b}^2 \Omega_{\rm Rb} = L_{\rm cen}
\ee
where $\Omega_{\rm Rb}= \Omega_{\rm 0}(R_{\rm b}/R_{\rm c})^{\beta}$ and,
for various values of $\beta$ and $\lambda$,
\be
\label{lambda}
\Lambda = \frac{2\left(3+\lambda\right)}{3\left(5+\lambda+\beta\right)}.
\ee
For example, for an initially uniform-density cloud in solid-body rotation,
$\Lambda$ = 2/5.
As in Bate \& Bonnell \shortcite{BatBon97}, we use natural units of 
$M_{\rm b}=1$ and $a=1$ for the `seed' binary, with $G=1$.

In general, to evolve the binary under the accretion of gas from 
the remainder of the collapsing cloud, we would need to know the 
infall rate and the specific angular momentum distribution 
of the gas as functions of time (i.e.~we would need to 
calculate how the cloud evolves as it collapses on to the binary).
However, if we make two simple assumptions, we can calculate
the evolution of the binary knowing only the density and angular 
momentum distributions of the cloud {\it before} it began to collapse.
Furthermore, we can consider the evolution of the binary as a 
function of the mass that has been accreted from the 
envelope $M_{\rm acc}$, and do not need to keep track of time explicitly.
First, we assume that the specific angular momentum of each element 
of gas in the cloud is conserved during its fall until it gets
very close to the binary.
Second, we assume that the time it takes for gas to fall on to the binary
from a radius $r$ from centre of the cloud is independent of direction
(i.e.~the gas falls on to the binary in spherical shells).

These are reasonable assumptions if the molecular cloud core 
undergoes a dynamic collapse (i.e.~the magnitude of its 
gravitational energy dominates the thermal, rotational and magnetic 
energies of the cloud).  In principle, angular momentum transport 
between elements of gas could occur during collapse if the 
cloud has density inhomogeneities (via gravitational torques)
or is threaded by magnetic fields (via magnetic torques).
However, if the collapse is dynamic, these effects are unlikely to have
enough time to transport a significant amount of angular momentum,
and even cores which are initially magnetically sub-critical 
typically evolve into configurations which undergo a dynamic 
collapse (e.g.~Basu 1997).

The integration of the binary from its initial mass, $M_{\rm acc}=M_{\rm b}=1$,
to its final mass $M_{\rm acc}=M_{\rm c} \geq M_{\rm b}$ 
(since some of the gas could be contained in a circumbinary disc)
under the accretion of the gas with $r>R_{\rm b}$ 
proceeds as follows (Figure \ref{model}).  
A spherical shell of gas with inner radius
$r=R_{\rm b}$ and thickness $\Delta r$ is divided into many
elements such that the gas in each element has the same specific
angular momentum (i.e. each `element' consists of two rings of 
gas, one above and one below the orbital plane).  The effect of
each element of gas on the binary, when it is accreted, is determined
using the results of Bate \& Bonnell \shortcite{BatBon97}
from its specific angular momentum, its mass, and $q$ (see below).  
The effects on the binary
from all the gas elements making up a shell are added together, and then the 
binary's properties ($M_{\rm b}$, $q$, $a$) are updated.  
The mass of material that is not captured by one of the two protostars
but that remains in a circumbinary disc (if any) is also determined.  
The process is
repeated for the shell of material at radius $r=R_{\rm b} + \Delta r$
until the entire cloud of mass $M_{\rm c}$ has been exhausted.

The results from Bate \& Bonnell \shortcite{BatBon97} are used to determine
the effects on the binary of the accretion of an element of gas.  
The specific angular momentum of an element of gas relative to the 
specific angular momentum required for gas
to form a circular orbit at a radius equal to the binary's separation 
is given by 
\be
\label{jrel}
j_{\rm rel} = \frac{r_{\rm xy}^2 \Omega_{\rm c}}{\sqrt{G M_{\rm b} a}}
\ee
where $r_{\rm xy}$ is the cylindrical radius of the gas element 
from the axis of rotation in the progenitor cloud.  To evolve the binary, we
need to know the values of
$\dot M_1/\dot M_{\rm acc}$, 
$\dot M_2/\dot M_{\rm acc}$ and 
$(\dot a/a)/(\dot M_{\rm acc}/M_{\rm b})$ as functions of $j_{\rm rel}$
and $q$,
where $\dot M_{\rm acc}$ is the rate of accretion from the infalling envelope
on to the binary.
Note that we consider gas to be `accreted' by one of 
the protostars if it is actually accreted onto the protostar itself or
if it is captured in the protostar's circumstellar disc 
(see also Section \ref{method}).
These quantities were determined by
Bate \& Bonnell \shortcite{BatBon97} only at set intervals in
$q$-$j_{\rm rel}$-space.  Interpolation between these points is performed 
to determine the quantities for any given $q$ and $j_{\rm rel}$.  
Boundary conditions are used for $q=1$ and $q=0$.  For $q=1$, we set
$\dot M_1/\dot M_{\rm acc} = \dot M_2/\dot M_{\rm acc}$
explicitly, with $(\dot a/a)/(\dot M_{\rm acc}/M_{\rm b})$ 
being determined 
from Bate \& Bonnell \shortcite{BatBon97}.  For $q=0$, we set 
$\dot M_{\rm 2}/\dot M_{\rm acc} = 0$ and 
\be
\begin{array}{l}
\dot M_{\rm 1}/\dot M_{\rm acc} = \dot M_{\rm b}/\dot M_{\rm acc} = \cases {1 & ~~~~~if $j_{\rm rel} \le 1$, \cr
                                        0 & ~~~~~otherwise \cr}
\end{array}
\ee
and $(\dot a/a)/(\dot M_{\rm acc}/M_{\rm b})$ is set equal to the 
values for $q=0.1$.  
The evolution of a binary with an initial mass ratio
$q \simgreat 0.05$ is insensitive to the assumptions for $q=0$. This
was tested by setting the primary and secondary accretion 
rates for $q=0$ equal to those for $q=0.1$ and repeating the calculations.
Note that we use the values of 
$(\dot a/a)/(\dot M_{\rm acc}/M_{\rm b})$
that were derived by Bate \& Bonnell \shortcite{BatBon97} 
for the effects of {\it accretion only}.  We do not include any
affect on the binary's separation due to the loss of orbital 
angular momentum to the gas in a circumbinary disc (if one exists).

\begin{figure}
\centerline{\hspace{0.0truecm}\psfig{figure=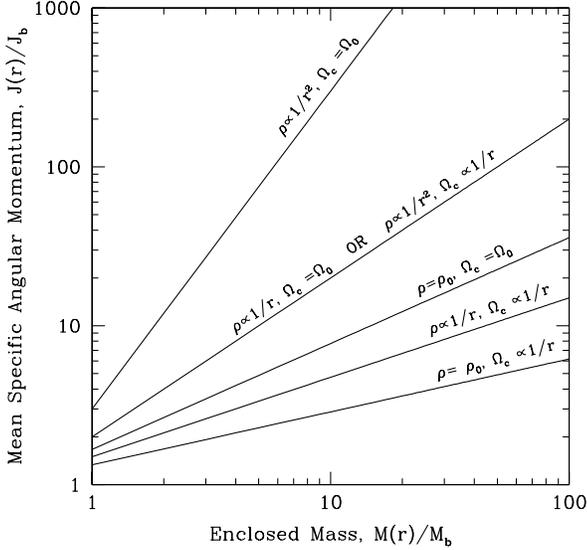,width=8.0truecm,height=8.0truecm,rwidth=8.0truecm,rheight=8.0truecm}}
\caption{\label{jrvsm}  The relationship between angular momentum and mass for different types of molecular cloud core (equation 7).  The values are normalised by the mean specific angular momentum and mass of the central region of gas from which the `seed' binary forms (which are also equal to the mean specific orbital angular momentum and mass of the `seed' binary itself).  Notice that the angular momentum is greater initially and increases more rapidly with enclosed mass in clouds which are more centrally condensed or have less differential rotation ($\beta \leq 0$).  Also, the relation between angular momentum and mass is the same for a cloud with $\rho \propto 1/r$ in solid-body rotation as it is for a cloud with $\rho \propto 1/r^2$ and $\Omega_{\rm c} \propto 1/r$.  }
\end{figure}

From equation \ref{lbinary}, we note that specifying the values 
of $\Lambda$ and $q$ fixes the relationship between the mean
specific angular momentum of a {\it shell} of gas, $J(r)$, at radius $r$,
and the enclosed mass, $M(r)$, in the progenitor cloud.
For all $q$,
\be
\label{jr}
\frac{J(r)}{J_{\rm b}} = \frac{2}{3\Lambda}\left( \frac{M(r)}{M_{\rm b}} \right)^{\left(\frac{2}{3\Lambda} - 1 \right)}
\ee
where $J_{\rm b}=L_{\rm b}/M_{\rm b}$ is the mean specific angular 
momentum of the `seed' binary, which depends on $q$.
Equation \ref{jr} is plotted for various types of molecular cloud core in
Figure \ref{jrvsm}.  With progenitor cores that are more centrally-condensed,
the specific angular momentum of the first gas to fall on to the 
`seed' binary from the envelope is greater and increases more rapidly
as mass is accreted than in less centrally-condensed cores.  
For cores which have a greater amount of differential rotation 
($\beta \leq 0$), the specific angular momentum of the gas is lower
to begin with and increases more slowly as mass is accreted.

Since $J_{\rm b}$ in equation \ref{jr} depends on the mass ratio, $q$, and
separation, $a$, of the `seed' binary, 
the angular momentum of the cloud, $J(r)$ is linked to the 
properties of the `seed' binary (due to our use of equation \ref{lbinary}).
This can be viewed in two ways.  First, if we consider a series 
of clouds with the same density and rotation profiles but that form
`seed' binaries with different mass ratios, then the separations
of the binaries will be somewhat larger for those with smaller 
mass ratios and $J_{\rm b}$ is the same for all mass ratios.
Alternately, if we choose `seed' binaries with the same 
separations, their progenitor clouds must be rotating more slowly
for those binaries with lower mass ratios.  We chose to use
equation \ref{lbinary} because, for a `seed' binary with a 
given mass ratio and separation, this gives the slowest possible
rotation rate of the progenitor cloud.  The progenitor cloud could
be rotating more rapidly than this if some of the angular momentum
of the gas from which the `seed' binary formed is contained in 
circumstellar discs, but it could not be rotating more slowly.

Finally, since choosing $\Lambda$ and $q$ fixes the relationship
between angular momentum and mass in the progenitor cloud
(independent of $M_{\rm c}/M_{\rm b}$, $R_{\rm c}$
and $\Omega_0$), it follows that the evolution proceeds 
in the same way for any value of $M_{\rm c}/M_{\rm b}$.  
Thus, a graph which shows the {\it evolution} of a particular
protobinary system as it accretes gas from its initial to its final
mass can {\it also} be viewed as giving the 
{\it final states} of binaries that have accreted a certain 
amount of mass relative to their initial mass (e.g.~Figures \ref{acc_d0_w0}
to \ref{acc_relax}).  This is only the case because we have chosen
clouds that have scale-free density and angular momentum profiles, 
and does not apply, for example, to clouds with Gaussian density
profiles that have a fixed inner to outer density contrast (e.g.~20:1).

\subsection{Smoothed particle hydrodynamics code}
\label{SPHcode}

To test the PBE code described above, we compare its results with those
obtained from full hydrodynamic calculations using a three-dimensional, 
smoothed particle hydrodynamics (SPH) code.  The SPH code is 
based on a version originally developed by Benz (Benz 1990; Benz et al.~1990).
The smoothing lengths of particles are variable in 
time and space, subject to the constraint that the number of neighbours
for each particle must remain approximately constant at $N_{\rm neigh}=50$.  
The SPH equations are integrated using a second-order Runge-Kutta-Fehlberg 
integrator with individual time steps for each particle (Bate et al.~1995).
Gravitational forces between particles and a particle's nearest neighbours 
are found by using either a binary tree, as in the original code, or the 
special-purpose GRAvity-piPE (GRAPE) hardware.  
The implementation of SPH using the GRAPE
closely follows that described by Steinmetz \shortcite{Steinmetz96}.
Using the GRAPE attached to a Sun workstation typically 
results in a factor of 5 improvement in speed over the workstation alone.

Calculations were performed using three different forms 
of artificial viscosity.  The scatter in the results from
calculations with different viscosities is used to give indication of the
uncertainty in the SPH results.  
The viscosity, $\Pi_{\rm ij}$, between 
particles $i$ and $j$ enters the momentum equation as
\be
\frac{{\rm d}{\bf v}_{\rm i}}{{\rm d}t} = - \sum_{\rm j}{ m_{\rm j} \left(\frac{P_{\rm i}}{\rho_{\rm i}^2} + \frac{P_{\rm j}}{\rho_{\rm j}^2} + \Pi_{\rm ij} \right) \nabla_{\rm i} W(r_{\rm ij}, h_{\rm ij})},
\ee
where ${\bf v}$ is the velocity, $t$ is the time, $m$ is the particle's mass,
$P$ is the pressure, $\rho$ is the density, $W$ is the SPH smoothing kernel, 
$r_{\rm ij}$ is the distance between particles $i$ and $j$, and $h_{\rm ij}$
is the mean of the smoothing lengths of particles $i$ and $j$.
All formulations of the viscosity have linear
and quadratic terms which are parameterised by $\alpha_{\rm v}$ and 
$\beta_{\rm v}$, respectively.  The first form of viscosity is the `standard'
form, originally given by Monaghan \& Gingold 
\shortcite{MonGin83} (see also Monaghan 1992)
\be
\Pi_{\rm ij} = \cases {\begin{array}{ll}
(-\alpha_{\rm v} c_{\rm ij} \mu_{\rm ij} + \beta_{\rm v} \mu_{\rm ij}^2)/\rho_{\rm ij} & {\bf v}_{\rm ij}\cdot {\bf r}_{\rm ij} \leq 0 \cr
0 & {\bf v}_{\rm ij}\cdot {\bf r}_{\rm ij} > 0 \cr
\end{array}}
\ee
where $\rho_{\rm ij} = (\rho_{\rm i} + \rho_{\rm j})/2$, $c_{\rm ij} = (c_{\rm i} + c_{\rm j})/2$ is the mean sound speed, ${\bf v}_{\rm ij} = {\bf v}_{\rm i} - {\bf v}_{\rm j}$, and
\be
\mu_{\rm ij} = \frac{h_{\rm ij} {\bf v}_{\rm ij}\cdot {\bf r}_{\rm ij}}{{\bf r}_{\rm ij}^2 + 0.01 h_{\rm ij}^2}.
\ee
This form is known to have a large 
shear viscosity. The second form (Monaghan \& Gingold 1983; 
Hernquist \& Katz 1989) 
has much
less shear viscosity but does not reproduce shocks quite as well as the 
`standard' formalism.  It is given by
\be
\Pi_{\rm ij} = \frac{q_{\rm i}}{\rho_{\rm i}} + \frac{q_{\rm j}}{\rho_{\rm j}}
\ee
where
\be
q_{\rm i} = \cases {\begin{array}{ll}
\alpha_{\rm v} h_{\rm i} \rho_{\rm i} c_{\rm i} |\nabla\cdot {\bf v}|_{\rm i} + \beta_{\rm v} h_{\rm i}^2 \rho_{\rm i} |\nabla\cdot {\bf v}|_{\rm i}^2 & (\nabla\cdot {\bf v})_{\rm i} \leq 0 \cr
0 & (\nabla\cdot {\bf v})_{\rm i} > 0 \cr
\end{array}}
\ee
The third form is that proposed by Balsara \shortcite{Balsara89} (see also
Benz 1990), which is
identical to the `standard' form, except that $\mu_{\rm ij}$ is replaced by 
$\mu_{\rm ij}(f_{\rm i}+f_{\rm j})/2$ where
\be
f_{\rm i}=\frac{|\nabla\cdot v|_{\rm i}}{|\nabla\cdot v|_{\rm i} + |\nabla\times v|_{\rm i} + \eta c_{\rm i}/h_{\rm i}}
\ee
and $\eta=1.0\times 10^{-4}$.
In purely compressional flows this form gives the same results as that of
the `standard' viscosity, 
while in shearing flows the magnitude of the viscosity 
is reduced.  The three forms of 
viscosity will be referred to as the Standard, $\nabla\cdot {\bf v}$, 
and Balsara forms, respectively.

Modelling of non-gaseous bodies (in this paper, the protostars), and the
accretion of gas on to them, is achieved by the inclusion of sink particles
(Bate et al.~1995; Bate 1995).  These were also used by Bate \& Bonnell
\shortcite{BatBon97}.  A sink particle is a non-gaseous particle
with a mass many times larger than that of an SPH gas particle.  Any SPH gas
particle that passes within a specified radius of the sink particle,
the accretion radius $r_{\rm acc}$, is accreted with its mass, linear 
momentum, and spin angular momentum being added to those of the sink particle.
Sink particles interact with SPH gas particles only via
gravitational forces.  Boundary conditions can also be included for particles
near the accretion radius (Bate et al.~1995), but they are not used for
the calculations in this paper.  Although boundary conditions are typically
required to stop the erosion of discs near the accretion radius 
(Bate et al.~1995), this is not necessary here because once a disc forms
it is replenished by the infalling gas more rapidly than it is eroded.

\section{Testing the Protobinary Evolution Code}
\label{comparison}

To test how accurately the protobinary evolution (PBE) code 
(Section \ref{PBEcode}) describes the evolution 
of a `seed' binary as it accretes from its initial to 
its final mass we performed several full SPH calculations for comparison.  
Two test cases were performed.  The first follows the formation of a 
binary system from the collapse of an initially uniform-density,
spherical molecular cloud core in solid-body rotation.  The 
`seed' binary is assumed to have a mass ratio of $q=0.6$ and a
mass of $M_{\rm b} = M_{\rm c}/10 = 1$.  The second test case
is similar, except 
that the progenitor cloud is centrally-condensed with a 1/r-density 
distribution
and there is less mass in the envelope relative to the initial binary's
mass: $M_{\rm b} = M_{\rm c}/5 = 1$.

\subsection{SPH initial conditions}

\subsubsection{Test Case 1}

The evolutions given by the PBE code are scale-free.  However, for 
the SPH calculations, we choose to specify physical parameters.  Test
case 1 consists of the collapse of a uniform-density
molecular cloud core in solid-body rotation with
\be
\begin{array}{ll}
M_{\rm c} = 1.0 ~{\rm M}_{\odot}, & R_{\rm c} = 1.0 \times 10^{17} ~{\rm cm}, \\
T = 10 ~{\rm K}, & \Omega_{\rm c} = 3.13 \times 10^{-14} ~{\rm rad~s}^{-1},
\end{array}
\ee
where $T$ is the temperature of the cloud, and it is assumed to consist of
molecular hydrogen.  The `seed' binary
is assumed to form from the mass initially contained within the sphere of 
radius $R_{\rm b} = 4.64 \times 10^{16}$ cm, which has 1/10 of the cloud's
total mass.

The initial conditions are evolved with $2.0 \times 10^5$ SPH gas particles
until the particles that were initially within $R_{\rm b}$ have collapsed
to within 
$r=2 \times 10^{15}$ cm of the centre.  At this point, the calculation is
stopped and the particles that were contained within $R_{\rm b}$ are removed
and replaced by a `seed' binary system with the same mass as those 
particles that were removed ($M_{\rm b}=0.1~{\rm M}_{\odot}$).  The binary
has mass ratio $q=0.6$, separation $a=1.0 \times 10^{15}$ cm and is in a 
circular orbit.  Note that the rotation rate of the cloud, $\Omega_{\rm c}$,
is chosen so that the total angular momentum of the gas initially within 
$R_{\rm b}$ is equal to the orbital angular momentum 
of the `seed' binary (the default used by the PBE code).  
The calculation is then
restarted and the remaining gas begins to fall on to the binary.
As the binary increases it mass by 5\%, its mass ratio and separation are
forced to evolve as predicted by the PBE code (to allow time for the
gas to establish its correct flow pattern on to the binary).  After
this, however, the binary is free to evolve as the accretion of gas dictates.

\subsubsection{Test Case 2}

The second test case consists of a 1/r-density
molecular cloud core in solid-body rotation with
\be
\begin{array}{ll}
M_{\rm c} = 1.0 ~{\rm M}_{\odot}, & R_{\rm c} = 1.0 \times 10^{17} ~{\rm cm}, \\
T = 10 ~{\rm K}, & \Omega_{\rm c} = 5.74 \times 10^{-14} ~{\rm rad~s}^{-1}.
\end{array}
\ee
The `seed' binary is assumed to form from the mass initially contained 
within the sphere of radius $R_{\rm b} = 4.47 \times 10^{16}$ cm, which
has 1/5 of the cloud's total mass.

The initial conditions are evolved with $1.0 \times 10^5$ SPH gas particles
(half the number that were used for test case 1 due to the smaller mass 
of the cloud
relative to the `seed' binary) in an identical manner to test case 1.

\subsection{Method}
\label{method}

While the gas is modelled using SPH particles in the usual manner,
the binary's components (the protostars) are modelled with 
sink particles (Section \ref{SPHcode}).  Their accretion
radii change in size as the mass ratio and separation of the 
binary changes and are always equal to 
$r_{\rm acc} = 0.1 R_{\rm i}$ where $R_{\rm i}$ are the 
sizes of the mean Roche lobes of the two protostars \cite{FraKinRai85}
\be
R_1/a = 0.38 - 0.20 ~{\rm log}~q, ~~~~~~~~~~~0.05 < q \leq 1,
\ee
for the primary and
\be
R_2/a = \cases{ \begin{array}{ll}
0.38 + 0.20 ~{\rm log}~q, & ~0.5 \leq q < 1, \cr
0.462~ (q/(1+q))^{1/3}, & ~~~\,0 < q < 0.5, \cr
\end{array}}
\ee
for the secondary.  This enables circumstellar discs of size 
$\simgreat 0.2 R_{\rm i}$ to be resolved by the SPH calculations.  Smaller 
accretion radii would allow smaller discs to be resolved, but would
also require more computational time due to the shorter dynamical
time-scales.

As in Bate \& Bonnell \shortcite{BatBon97}, we define the mass of a protostar
($M_1$ or $M_2$) to be the mass of a sink particle {\it and its circumstellar 
disc} (if any).  Hence, when stating the masses of the two components or 
mass ratio of the binary we are assuming that, in reality, all of the 
material in a circumstellar disc will eventually be accreted by its
central star.
Therefore, the binary's mass $M_{\rm b}$ is equal to the masses of the sink
particles and their circumstellar discs.  The parameters of the binary's 
orbit are calculated by considering these masses.

A gas particle belongs to a circumstellar disc if
its orbit, calculated considering only one sink particle at a time, has
eccentricity $<0.5$, and semi-major axis $<D/2$.  This simple 
criterion gives excellent extraction of the circumstellar discs.
The mass of material in a circumbinary disc $M_{\rm cb}$ (if any),
is defined as being the gas which has an outwards (positive) radial
velocity or which is falling on to the binary more slowly than 1/3 of the 
local free-fall velocity (i.e.~$v_{\rm r} > -1/3 \sqrt{2 G M_{\rm b}/r}$) 
and which is not in one of the
circumstellar discs as defined above or within distance $a$ of the
binary's centre of mass.  The amount of gas that has fallen on to the 
binary from the envelope $M_{\rm acc}$ is defined as $M_{\rm b}+M_{\rm cb}$
plus the remaining gas within $a$ of the binary's centre of mass.

\begin{figure*}
\vspace{-0.7truecm}
\centerline{\hspace{0.2truecm}\psfig{figure=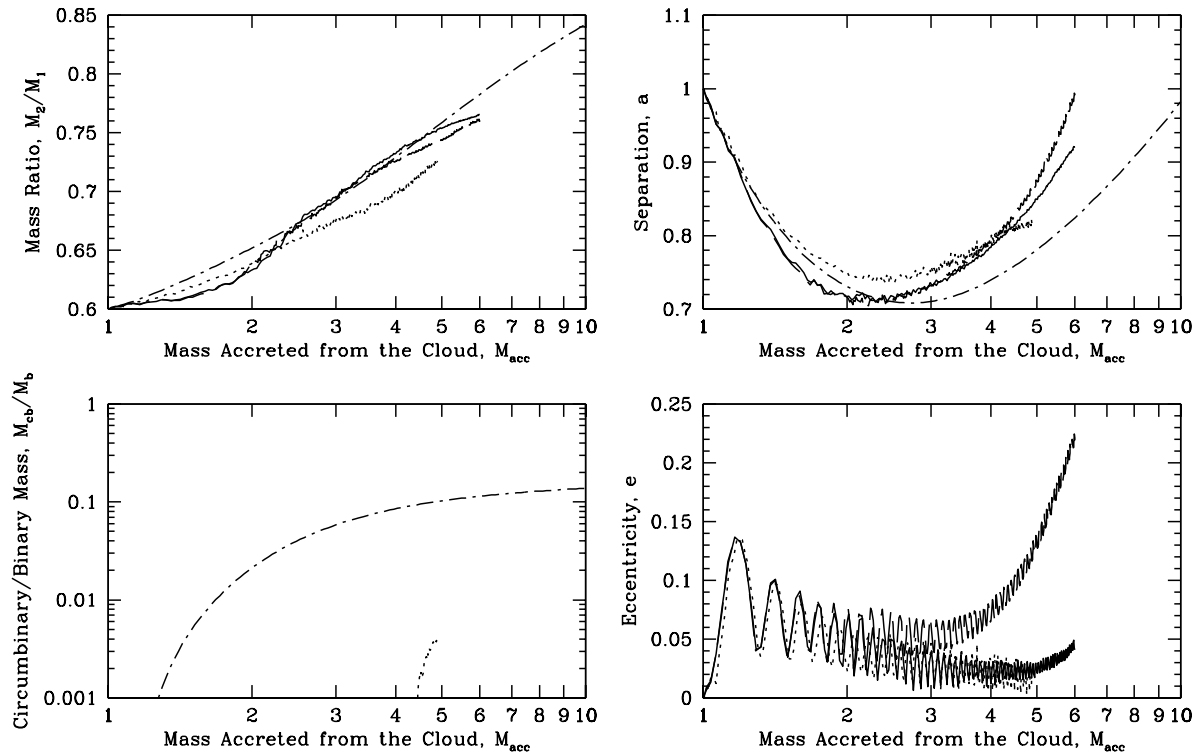,width=17.0truecm,height=17.0truecm,rwidth=17.0truecm,rheight=11.0truecm}}
\caption{\label{PBESPH1}  First comparison of the results from the protobinary evolution (PBE) code with those from the full SPH calculations.  The graphs show the evolution of a protobinary system that was formed in the centre of a collapsing molecular cloud core which initially had a uniform density profile and was in solid-body rotation.  The evolution of the (a; upper left) mass ratio, (b; upper right) separation, (c; lower left) ratio of mass in a circumbinary disc to that of the binary, and (d; lower right) eccentricity are plotted as the binary accretes gas from the infalling envelope.  The dot-dashed lines show results from the PBE code.  The other lines show the results from the full SPH calculations with artificial viscosities of S1 (long-dashed), S2 (solid), and B1 (dotted) (see Section 3.3.2).}
\end{figure*}

The equation of state of the gas is given by
\be
P = K \rho^{\gamma},
\ee
where $P$ is the pressure, $\rho$ is the density, and $K$ is a constant 
that depends on the entropy of the gas.  The ratio of specific heats $\gamma$
varies with density as
\be
\gamma = \cases{\begin{array}{ll}
1.0, & \rho \leq 10^{\eta}~ {\rm g~cm}^{-3}, \cr
1.4, & \rho > 10^{\eta}~ {\rm g~cm}^{-3}. \cr
\end{array}}
\ee
The value of $K$ is defined such that when the gas is 
isothermal $K=c_{\rm s}^2$, with 
$c_{\rm s} = 2.0 \times 10^4$ cm s$^{-1}$ for molecular hydrogen at 10 K.  
We choose $\eta=-12$ for test case 1 and $\eta=-13$
for test case 2.  The heating of the gas at high densities inhibits
fragmentation of the discs
(e.g.~Bonnell 1994; Bonnell \& Bate 1994a; Burkert \& Bodenheimer 1996; 
Burkert et al.~1997).  
In reality, this heating occurs when the gas becomes optically thick to
infrared radiation.  The results are independent of the exact
density at which heating starts, so long as the gas which falls on to the
binary from the envelope is `cold' (i.e.~pressure forces are not dynamically
important so that the gas falls freely on to the binary), 
the discs are relatively thin, and the discs do not fragment.

\subsection{Test Case 1}
\label{testcase1}

The evolutions of test case 1 as given by the PBE and SPH codes are
presented in Figure \ref{PBESPH1}.  We give the binary's mass ratio $q$,
separation $a$, the ratio of the mass of the circumbinary disc (if one exists)
to the mass of the binary $M_{\rm cb}/M_{\rm b}$, and the binary's eccentricity
$e$, as functions of the mass which has been accreted from 
the cloud $M_{\rm acc}$ (up to 10 times the binary's initial mass).
In addition, for two of the SPH calculations, we show snapshots at various
times during the calculations (Figures \ref{UD_S2} and \ref{UD_B1}).

\subsubsection{PBE calculation}

The PBE code predicts that after the binary has accreted all the gas in
the cloud, its mass ratio has increased from $q=0.6$
to $q \approx 0.84$ while its separation, after decreasing initially, has
returned to approximately the initial value.  The binary's final 
mass is 8.8 times its initial mass
with the remaining 12\% of the cloud 
located in a circumbinary disc.  With the PBE
code, the binary is assumed to remain in a circular orbit throughout the
evolution.

\subsubsection{SPH calculations}

Three SPH calculations were performed using different formulations of
the artificial viscosity: 
S1: Standard, with $\alpha_{\rm v}=1$ and $\beta_{\rm v}=2$;
S2: Standard, with $\alpha_{\rm v}=0.5$ and $\beta_{\rm v}=2$; and B1: Balsara,
with $\alpha_{\rm v}=0.25$ and $\beta_{\rm v}=1$.  S1 and S2 have the same
formulation, but S2 has approximately half the shear viscosity that is
present in S1 (the linear $\alpha_{\rm v}$-viscosity dominates the shear
viscosity in the code; the $\beta_{\rm v}$-viscosity is only important
in shocks).  B1 has the lowest shear
viscosity of the three, but also has a different formulation.

Unfortunately, the SPH calculations cannot be evolved until all of the
original cloud has been accreted.  The reason is that as the 
evolution proceeds,
the rate at which mass falls on to the binary decreases and, thus,
more orbits of the binary 
must be calculated for the same amount of mass to fall on to the binary.
For example, the increase from $M_{\rm b}=1$ to $M_{\rm b}=2$ takes $\approx 5$
orbits, while the increase from $M_{\rm b}=4$ to $M_{\rm b}=5$ takes 
$\approx 14$
orbits.  The CPU time required to evolve the binary until the entire cloud
falls in is prohibitive, which, after all, is the reason that we
developed the PBE code in the first place.  It takes $\approx 60$ orbits
for the binary to increase its mass by a factor of 6 (i.e.~$\approx 60$\% 
of the total cloud was accreted).
Each of the three SPH calculations took $4-5$ months on 
a 170 MHz Sun Ultra workstation with a GRAPE board.  
The evolution with the PBE code takes a few seconds!

\subsubsection{Evolution of the mass ratio and separation}

Although the SPH calculations do not run to completion, 
we can compare the evolution as the binary's mass increases 
by a factor of 5-6.  Overall, there is excellent agreement 
between the PBE and SPH codes for the evolution of the mass ratio
and separation.  
The mass ratio is predicted to within 5\% over the entire evolution
and the separation to within 20\%.  Indeed, there is as much 
scatter between the different SPH results 
as there is between the PBE results and the SPH results.  
Thus, {\bf for the evolution of the mass ratio
and separation, we conclude that the PBE code is at least
as accurate as a full SPH calculation of this resolution.}
Notice also that the PBE results (which assumed $e=0$) and SPH 
results are in good agreement even though the eccentricity varies 
between $0<e\simless 0.2$ during the SPH calculations.  This implies
that the evolution given by the PBE code is satisfactory, not just for
circular binaries, but for binaries with $e \simless 0.2$.

\begin{figure*}
\vspace{7.00truecm}
\caption{\label{UD_S2}  Snapshots of the evolution of test case 1 using the SPH code with artificial viscosity S2.  The panels show the logarithm of column density, looking down the rotation axis, as the binary accretes infalling gas.  The primary is on the right.  The calculation starts when the mass accreted from the cloud is $M_{\rm acc}=1$ and is followed until $M_{\rm acc}=6$.  Notice that, initially, the circumsecondary disc is too small to be resolved and it only begins to be resolved when $M_{\rm acc}\simgreat 2.0$.  Each panel has a width of twice the binary's initial separation.  Snapshots from calculations S1 and S2 look almost identical.}
\end{figure*}

\begin{figure*}
\vspace{7.00truecm}
\caption{\label{UD_B1}  The same as in Figure 4, except using artificial viscosity B1 and only following the binary until $M_{\rm acc}=5.0$.  With the different viscosity, the circumsecondary disc is unresolved for much longer (until $M_{\rm acc}\simgreat 4.0$) and a circumbinary disc starts to form earlier at $M_{\rm acc}\approx 4.4$.  Each panel has a width of 4 times the binary's initial separation. }
\end{figure*}

\subsubsection{Evolution of the circumbinary disc}

The good agreement for the mass ratio and separation does not 
hold for the evolution of the circumbinary disc.  The 
PBE code predicts that more gas will settle into a circumbinary 
disc than is found in the SPH calculations.  
Cases S1 and S2 do not form circumbinary discs 
at all for $M_{\rm acc}<6$, while B1 only begins to form 
a circumbinary disc when $M_{\rm b}\simgreat 4.3$ 
(Figures \ref{PBESPH1}c and \ref{UD_B1}).  The difference between
S1/S2 and B1 can be attributed to the much larger shear viscosity which
is present in S1 and S2 compared to B1.  Greater shear viscosity inhibits 
the formation of a circumbinary disc by removing angular momentum 
from the gas which would otherwise settle into a circumbinary disc.
Another problem is that, in all of the SPH calculations, the binary
has a larger separation than is predicted by the PBE code when
$M_{\rm acc}\simgreat 2.3$ which also inhibits the formation of
a circumbinary disc.  These problems of circumbinary disc formation
are the main reason that we performed test case 2
and we defer further discussion to Section \ref{testcase2}.

\subsubsection{Dependence of the results on the circumstellar discs}

Returning to the evolution of the mass ratio and separation, 
although the overall agreement is good, there are two
differences which need, briefly, to be commented on.
These differences are related to the way in which the PBE and 
SPH codes handle the circumstellar discs.

First, S1 and S2 both give a slower rate of increase of the 
mass ratio initially than is predicted by the PBE code 
(Figure \ref{PBESPH1}a; $1 \leq M_{\rm acc}
\simless 2$).  This reflects a problem with the SPH calculations rather than
with the PBE code: namely, the size of the accretion radius 
around the secondary is initially of a similar size to that of the 
circumstellar disc which should form.  This stops a circumsecondary disc 
from forming which reduces the secondary's cross-section 
for capturing infalling gas and, so, reduces the amount of gas captured
by the secondary.  Only when $M_{\rm acc}\simgreat 2.0$, is
the specific angular momentum of the gas being captured by the secondary
large enough for it to form a circumsecondary disc outside the accretion
radius (Figure \ref{UD_S2}).  From this point on, the mass 
ratio increases at the rate predicted by the PBE code.  
B1 has a similar problem, but for a slightly different reason.  B1
has much less shear {\it and bulk} viscosity in shearing flows.  In order
for infalling gas to form a resolved disc around the secondary, it
must, first, dissipate most of its kinetic energy so that it is captured
by the secondary and, second, dissipate enough kinetic energy that the
gas particles have roughly circular orbits.  If the gas does not dissipate
enough kinetic energy to be captured by the secondary on its first pass, 
it is likely to be captured by the primary instead which is deeper 
in the gravitational potential well.  If a gas particle is captured by the
secondary but still has a very elliptical orbit, it will pass inside the
secondary's accretion radius and be removed.  Thus, the low bulk 
viscosity of B1 means that the formation of a resolved 
circumstellar disc around the secondary is delayed until 
$M_{\rm acc}\simgreat 3.9$ (Figure \ref{UD_B1}) and the 
cross-section of the secondary is underestimated until this point.
As with S1 and S2, however, as soon as the circumsecondary disc is 
resolved ($M_{\rm acc}\simgreat 3.9$), the mass ratio begins to
increase at the rate predicted by the PBE code (Figure \ref{PBESPH1}a).

Second, the PBE code does not include viscous 
evolution of the circumstellar discs.  In reality, 
and in the SPH calculations, such viscous
evolution results in the transfer of angular momentum from the 
discs to the orbit of the binary, increasing the separation.

In order for us to be able to neglect viscous evolution of the 
circumstellar discs, this transfer of angular momentum must be negligible 
over the time for most of the envelope to be accreted.  Viscous
evolution of the circumstellar discs on a longer time-scale may 
increase the final separation of the binary by a small factor, 
but the masses of the binary's components will have been
determined during the accretion phase.

For wide binaries, viscous evolution is unlikely to be significant.
For example, the free-fall time for a dense molecular cloud core 
($10^{-18}$ g cm$^{-3}$) is $\sim 10^5$ years.  The envelope
will fall on to the binary on this time-scale.  The viscous time for
one of the circumstellar discs is of order 
\be
t_{\rm v} \sim \frac{P}{2 \pi \alpha_{SS}}\left(\frac{R}{H}\right)^2,
\ee
where $P$ is the period of the binary, $\alpha_{SS}$ measures the strength of
the shear viscosity using the standard Shakura--Sunyaev prescription, and
$H/R$ is the ratio of the thickness of the disc to its radius (typically
$\sim 0.1$; Burrows et al.~1996; Stapelfeldt et al.~1998).  Observationally,
it is thought that the shear viscosity acting in protostellar accretion
discs is of order $\alpha_{\rm ss} \sim 0.01$ \cite{Hartmannetal98}.  
Thus, taking the median 
binary separation of 30 AU \cite{DuqMay91} and a typical protobinary mass
of $0.1 {\rm M}_\odot$, the viscous time in one of the circumstellar 
discs is $\sim 10^6$ years, an order of magnitude longer
than the free-fall time.  For close binaries
(separations $\simless 5$ AU) viscous evolution may be expected to
have some effect during the accretion phase.  However, in most cases, 
a close binary is expected to have a circumbinary disc 
(see Section \ref{circumbinary})
and interactions between it and the binary will dominate the 
interactions between the circumstellar discs and the binary 
(Section \ref{testcase2cb}).

In SPH calculations S1 and S2, viscous evolution of the circumstellar discs
does affect the binary's separation during the calculation.
For example, in calculation S1 when $M_{\rm acc}\simgreat 3$, 
the separation increases
more rapidly than predicted by the PBE code and when $M_{\rm acc}=6$,
after $\approx 60$ orbits ($\approx 4 \times 10^4$ years) the separation 
is 20\% larger than predicted.  However,  
we estimate (Pongracic 1988; Meglicki, Wickramasinghe \& Bicknell 1993;
Artymowicz \& Lubow 1994)
the effective viscosity in the circumstellar discs 
to be $\alpha_{\rm SS}\approx 0.2$.
With an orbital period of $\approx 600$ years over most of the evolution, 
this gives a viscous time of 
$\sim 5 \times 10^4$ years which is similar to the time over which 
the calculation was followed.  Therefore, 
it is not surprising that the separation was affected.  We note,
however, that this viscosity is $\approx 20$ times stronger than is
expected in real protostellar discs and, thus, the effect on a real
protobinary system over this period would have been negligible,
as is assumed by the PBE code.  Calculation S2 has approximately
half the shear viscosity of S1 and its separation is correspondingly
closer to the 
PBE code prediction.  B1 has much less shear viscosity than S1 or S2.
Consequently, the {\it rate of change} of separation when 
$M_{\rm acc}\simgreat 3$ is very close to that predicted by the PBE 
code.  Instead, with B1, the differences in the evolution of the 
separation occurred earlier in the evolution 
($1.5\simless M_{\rm acc} \simless 2.5$).

We conclude that it is reasonable for the PBE code to neglect the 
effect on the separation due to viscous evolution of the 
circumstellar discs
since it is only likely to have a significant effect for close 
binaries and in these cases the effect is likely to be overwhelmed 
by the interaction between the binary and a circumbinary disc.  
We also find that the small differences between the PBE and SPH results 
for the evolution of the mass ratio and separation reflect 
the unphysical treatment of the circumstellar 
discs by the SPH code rather than a problem with the PBE code.

\begin{figure*}
\vspace{-0.7truecm}
\centerline{\hspace{0.2truecm}\psfig{figure=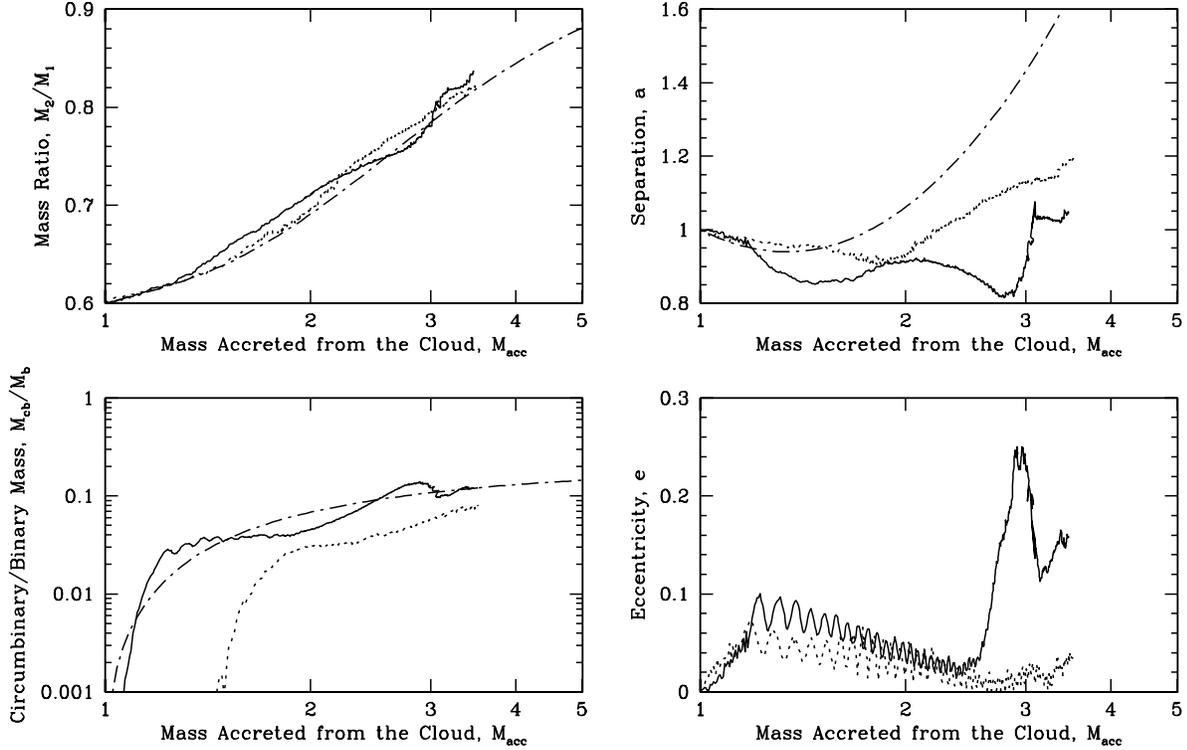,width=17.0truecm,height=17.0truecm,rwidth=17.0truecm,rheight=11.0truecm}}
\caption{\label{PBESPH2}  Second comparison of the results from the protobinary evolution (PBE) code with those from the full SPH calculations.  The graphs show the evolution of a protobinary system that was formed in the centre of a collapsing molecular cloud core which initially had a $1/r$-density profile and was in solid-body rotation.  The evolution of the (a; upper left) mass ratio, (b; upper right) separation, (c; lower left) ratio of mass in a circumbinary disc to that of the binary, and (d; lower right) eccentricity are plotted as the binary accretes gas from the infalling envelope.  The dot-dashed lines show results from the PBE code.  The other lines show the results from the full SPH calculations with artificial viscosities of B1 (dotted) and D1 (solid) (see Section 3.4.2).}
\end{figure*}

\subsubsection{Eccentricity evolution}

Finally, we comment on the eccentricity of the binary as given by the
SPH code.  
We see that initially, independent of the viscosity, the binary
becomes eccentric due to the accretion of gas, with a peak eccentricity of
$e \approx 0.14$.  The eccentricity then displays a secular decrease, 
as the accretion rate on to the binary also decreases,
while oscillating
on an orbital time-scale.  Such periodic changes of the orbital
elements are known from analytic solutions of binary motion under a
perturbing force (Kopal 1959), in this case, the accreting gas.  
Later in the evolution,
the eccentricity displays a secular increase which depends on the 
shear viscosity in the calculations: the greater the shear viscosity,
the earlier the increase begins.  As with the overestimate of the
rate of change of orbital separation that was given by S1 and S2 when 
$M_{\rm acc}\simgreat 3$, the eccentricity increases due to the 
unphysically-rapid transfer of angular momentum and energy from 
the circumstellar discs into the binary's orbit.

\subsection{Test case 2}
\label{testcase2}

The second test case was chosen primarily to study the differences
that appeared in test case 1 between the PBE and SPH codes regarding
the formation of a circumbinary disc.  A more centrally-condensed initial
cloud results in the gas which first falls on to the binary having
more specific angular momentum than with a uniform-density 
cloud and in a more rapid increase of the specific angular momentum of
the gas as the accretion proceeds (see Section \ref{soliddenr}).  
Therefore, a circumbinary disc should form earlier than with
test case 1 and provide us with a better test for the evolution of the 
circumbinary disc.
The evolutions given by the PBE code and two different SPH calculations 
are given in Figure \ref{PBESPH2}, with snapshots from the SPH calculations
in Figures \ref{1r_B1} and \ref{1r_D1}.

\subsubsection{PBE calculation}

The PBE code predicts that at the end of the
evolution the mass ratio should have increased from $q=0.6$
to $q \approx 0.88$ while the separation, after decreasing slightly
at the beginning, finally ends up at $\approx 2.3$ times the initial
separation.  The mass of the circumbinary disc becomes significant
($M_{\rm cb}/M_{\rm b} \simgreat 1/20$) at $M_{\rm acc} \approx 1.5$,
and at the end of the calculation the circumbinary disc contains
$\approx 13$\% of the total mass.

\subsubsection{SPH calculations} 

The SPH calculations had two different forms of artificial viscosity: 
B1: Balsara, with $\alpha_{\rm v}=0.25$ and $\beta_{\rm v}=1$;
D1: $\nabla\cdot {\bf v}$, also with $\alpha_{\rm v}=0.25$ and 
$\beta_{\rm v}=1$.  Both formulations have low shear viscosity.  The 
bulk viscosities are similar in non-shearing flows, but in shearing 
flows B1 has less bulk viscosity.  We do
not use the Standard viscosity formulation (S1 or S2 in test case 1)
because of our finding that its large shear viscosity inhibits 
the formation of a circumbinary disc and leads to rapid evolution of 
the circumstellar discs.

As with test case 1, due to the computational cost, the SPH 
calculations are stopped before all of the gas has fallen on to the binary.
Each calculation took $\approx 4$ months on a 300 MHz Sun Ultra 
workstation (using the binary tree, not a GRAPE board).  During the
calculations, the binaries performed $\approx 60$ orbits and $\approx 70$\%
of the total mass was accreted by the binary or settled into a circumbinary
disc.

\begin{figure*}
\vspace{3.6truecm}
\caption{\label{1r_B1}  Snapshots of the evolution of test case 2 using the SPH code with artificial viscosity B1.  The panels show the logarithm of column density, looking down the rotation axis, as the binary accretes infalling gas.  The primary is on the right.  The evolution starts when the mass accreted from the cloud is $M_{\rm acc}=1$ and is followed until $M_{\rm acc}=3.5$.  Two circumstellar discs are formed as soon as the calculation begins and a circumbinary disc begins to form at $M_{\rm acc}\simgreat 1.4$.  Each panel has a width of 8 times the binary's initial separation. }
\end{figure*}

\begin{figure*}
\vspace{3.6truecm}
\caption{\label{1r_D1}  The same as in Figure 7, except using artificial viscosity D1.  With the different viscosity, the circumbinary disc forms earlier (at $M_{\rm acc}\approx 1.1$) and the spiral shocks in the disc are better resolved.}
\end{figure*}

\subsubsection{Evolution of the mass ratio} 

The agreement for the evolution of the mass ratio is even better than it
was with test case 1 with differences of $\simless 3$\%.  This
is because, in this test case, the specific angular momentum 
of the gas being 
captured by the secondary is large enough for it to form a 
circumsecondary disc outside the accretion radius as soon as the 
SPH calculations begin, regardless of the viscosity 
(Figures \ref{1r_B1} and \ref{1r_D1}).  The SPH calculations
generally give a slightly higher mass ratio than is predicted by the
PBE code which is exactly what is expected because the PBE code 
ignores the separation-decreasing effect of the interaction between
the binary and the circumbinary disc (a smaller binary separation means
that the specific angular momentum of the gas is higher relative to
that of the binary, resulting in more gas being captured by the 
secondary).

\subsubsection{Circumbinary disc evolution and interactions} 
\label{testcase2cb}

For test case 2, the differences between the PBE and SPH results involve the
formation of the circumbinary disc and its interaction with the binary,
the latter of which is not accounted for in the PBE code.  
Generally, a binary will interact with a circumbinary disc
so as to transfer angular momentum from its orbit into the gas of
the circumbinary disc.  This tends to decrease the separation of the
binary and increase its eccentricity (Artymowicz et al.~1991; Lubow
\& Artymowicz 1996).  
Both these effects can be seen in the SPH calculations.  The separation
follows the prediction of the PBE code to better than $3$\%
until the circumbinary disc begins to form, after which it is 
always smaller than predicted by the PBE code.  As in test case 1, 
the eccentricity grows at the start of the
calculation and then displays a secular decrease with time.  However,
when the mass of the circumbinary disc exceeds $\approx 0.05 M_{\rm b}$,
the transfer of angular momentum from the binary to the circumbinary disc
overwhelms the eccentricity-decreasing effect of the infalling gas and
the eccentricity increases.

As with test case 1, different formulations of the SPH viscosity give
slightly different evolutions.  
D1 results in earlier formation of a circumbinary disc than B1, although
at $M_{\rm acc}=2$ and $M_{\rm acc}=3.5$ 
the circumbinary disc masses only differ by 
$\approx 30$\%.  More importantly, we find that B1 gives very poor 
resolution of spiral shocks in the circumbinary disc (created by 
gravitational torques from the binary) due to its lower bulk viscosity
in shearing flows 
compared to D1 (c.f.~Figures \ref{1r_B1} and \ref{1r_D1}).
The better shock resolution of D1
results in more efficient angular momentum transport from the binary's
orbit to the circumbinary disc which leads to
a smaller binary separation and, consequently, more of the infalling gas
settles into the circumbinary disc than with B1.  This can be seen in
the rapid decrease in separation of D1 compared to B1 when 
$2.0 \simless M_{\rm acc}\simless 2.6$ (Figure \ref{PBESPH2}b), and, 
simultaneously, the more rapid 
increase in the mass of the circumbinary disc for D1 (Figure \ref{PBESPH2}c).  

From the point of view of the ability to resolve shocks in the 
circumbinary disc, D1 is more realistic than B1
and we emphasise that extreme care should be used when employing 
the Balsara formulation of viscosity in shearing flows.  Given that
D1 appears to give more realistic results than B1, is it encouraging to
note that over the entire evolution the PBE code predicts the 
circumbinary disc mass to within a factor of 2 of that given by the D1 SPH
calculation (Figure \ref{PBESPH2}).

\begin{figure*}
\vspace{-0.7truecm}
\centerline{\hspace{0.2truecm}\psfig{figure=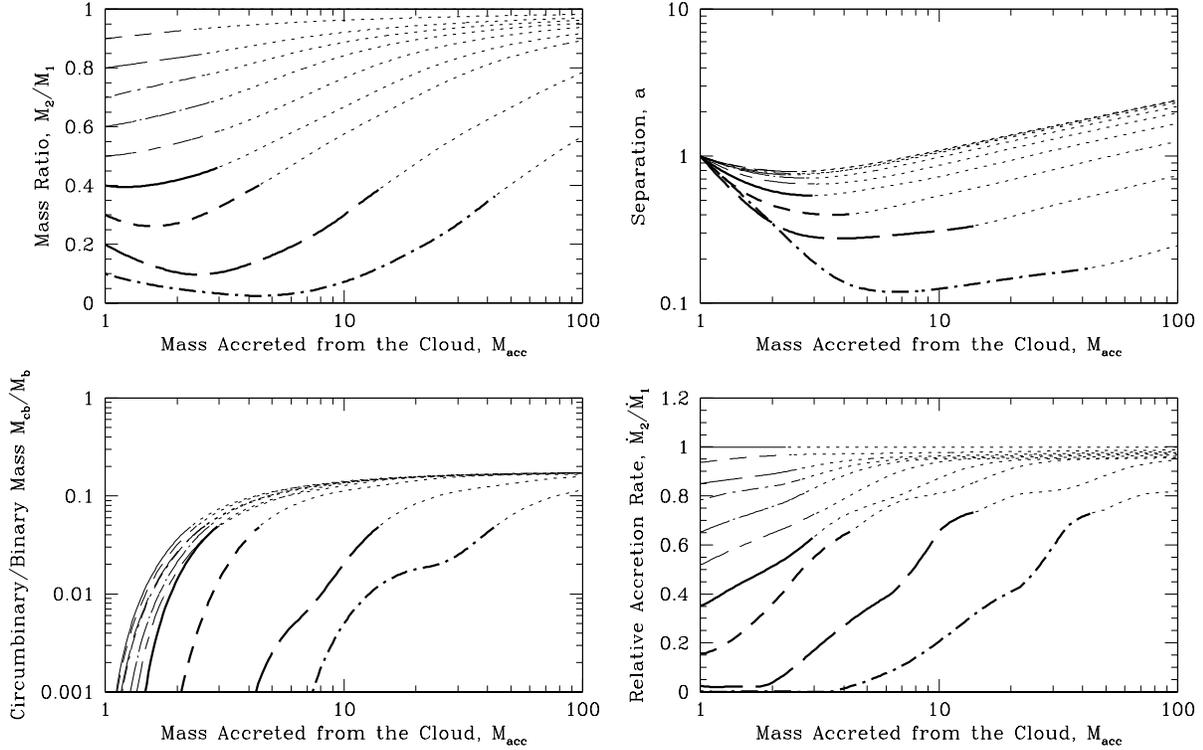,width=17.0truecm,height=17.0truecm,rwidth=17.0truecm,rheight=11.0truecm}}
\caption{\label{acc_d0_w0}  The evolution of protobinary systems which were formed in the centres of collapsing molecular cloud cores as they accrete from their gaseous envelopes.  The initial cores have uniform-density profiles and are in solid-body rotation.  The evolution of the mass ratio (a; upper left), separation (b; upper right), and ratio of the circumbinary disc mass to that of the binary (c; lower left), and the relative accretion rate (d; lower right) are given as functions of the amount of gas that has been accreted from the envelope.  The evolutions are given for binary systems with initial mass ratios of $q=0.1$ to $q=1.0$, with various different line types and/or line widths for each.  The functions are all given by thin dotted lines once the circumbinary disc mass exceeds 5\% of the binary's mass.  Beyond this point, the binary's mass ratio and the mass in the circumbinary disc are likely to be underestimated, and the separation is likely to be overestimated.  Masses are given in units of the binary's initial mass; separation is given in units of the binary's initial separation.  }
\end{figure*}

\subsection{Conclusions and limitations of the PBE code}

In summary, in the absence of a massive circumbinary disc, 
the PBE code gives results which are in good qualitative
and quantitative agreement with the full hydrodynamic calculations.
In all cases, the binary's mass ratio is predicted to within 5\% of
the SPH results {\it over an increase in the binary's total mass 
by a factor of up to 6! }  In test case 1, the binary's separation 
is predicted to within 20\% by the PBE code.  Furthermore, in test
case 1, even these small differences can be attributed to the 
SPH code's inability to resolve the circumsecondary disc during 
parts of the calculations or unphysically-rapid viscous evolution
of the circumstellar discs, rather than a problem with the PBE code.
Thus, {\bf the results from the PBE code are at least as accurate as 
those given by a full SPH calculation, but the 
PBE code is $\sim 10^6$ times faster}.  
This makes it possible to investigate the
statistical properties of binary stars (see the next section).

In cases where a circumbinary disc forms around the binary,
the PBE code generally predicts that the 
disc forms earlier than in the full SPH calculations, especially in the
more borderline case of 
a progenitor molecular cloud core with uniform-density.  However, in the
best case for studying the formation of a circumbinary disc 
(test case 2, D1) the time of formation of the circumbinary disc was
in good agreement and its mass was predicted to within a factor of 2
throughout the evolution.

The main omission in the PBE code is that, if a massive circumbinary disc
is formed ($M_{\rm cb}/M_{\rm b} \simgreat 1/20$), the code ignores 
the interaction between it and the binary.  Omitting this interaction
leads to the separation of the binary being {\it overestimated} by the 
PBE code.  From the SPH results, we find that {\bf if a massive 
circumbinary disc is formed, the binary's separation is likely 
to remain approximately constant as it accretes from a gaseous envelope}.  
The overestimate of the separation by the PBE code means that
{\bf the mass of the circumbinary disc is likely to be underestimated and,
for the same increase in the binary's mass, the binary's mass ratio will be
underestimated}.  We note, however, that this omission serves 
only to {\it strengthen} the predictions of the properties 
of binary stars that we obtain in Section \ref{predictions}.

\begin{figure*}
\vspace{-0.7truecm}
\centerline{\hspace{0.2truecm}\psfig{figure=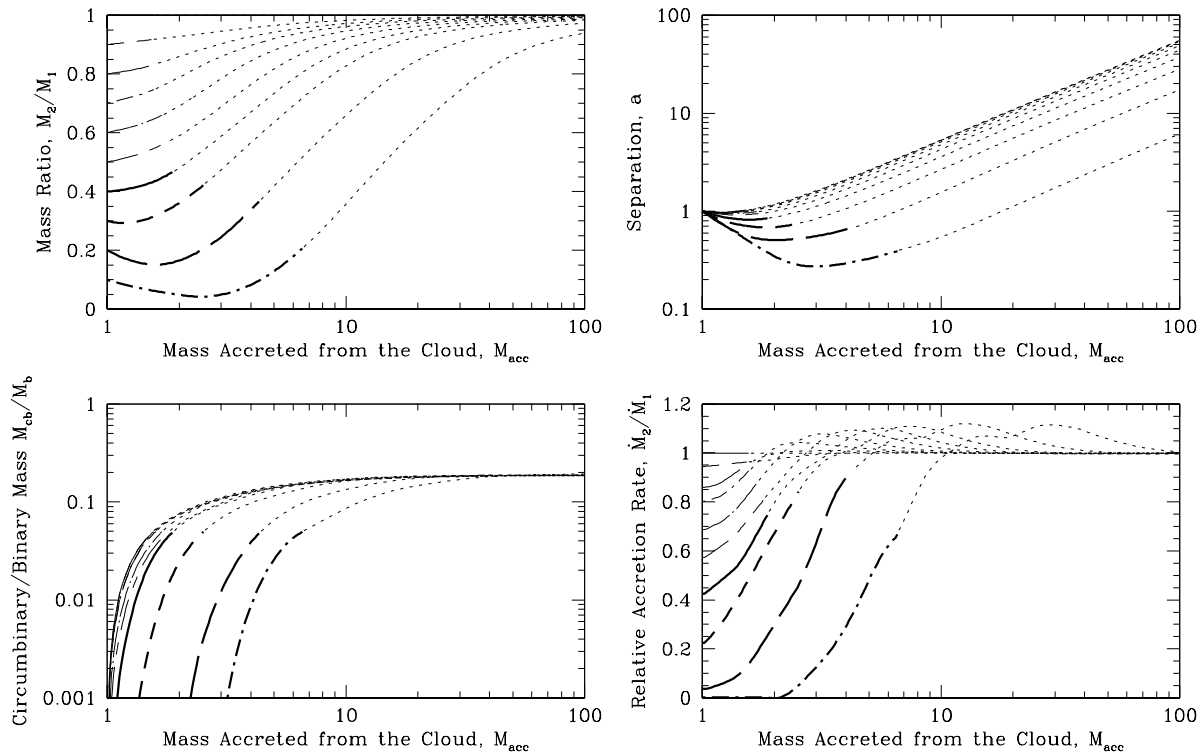,width=17.0truecm,height=17.0truecm,rwidth=17.0truecm,rheight=11.0truecm}}
\caption{\label{acc_d1_w0}  The evolution of protobinary systems which were formed in the centres of collapsing molecular cloud cores as they accrete from their gaseous envelopes.  The initial cores have {\it either} $1/r$-density profiles and are in solid-body rotation, {\it or} $1/r^2$-density profiles and are in differential rotation with $\Omega_{\rm c} \propto 1/r$.  These two types of core have the same relationship between angular momentum and mass (see Figure 2).  See Figure 9 for a description of the plots.  }
\end{figure*}

\section{Protobinary evolution}
\label{evolution}

We now use the PBE code to study the 
evolution and final states of binaries that are formed 
from the collapse of 6 types of molecular cloud core.  We study 
binaries that are formed from clouds with three different initial density
profiles: uniform-density, and power-law profiles of $\rho \propto 1/r$
and $\rho \propto 1/r^2$ (i.e. $\lambda$=0, -1, and -2).  
For each of these three density profiles, we study two different initial
rotation profiles: solid-body rotation, 
and $\Omega_{\rm c} \propto 1/r$ (i.e. $\beta$=0, and -1).

These initial conditions are highly idealised.  However, we use them
to illustrate how the properties of a binary depend on the 
degree of central condensation and the amount of differential rotation
in its progenitor molecular cloud core.  In fact, in
several cases the initial conditions represent extremes; the initial conditions
before dynamic collapse occurs are almost certainly somewhat 
centrally-condensed, but are almost certainly less centrally condensed than
$\rho \propto 1/r^2$.  In Section \ref{predictions}, we use the conclusions
reached from these calculations to predict the properties of binaries
and to constrain the properties of the molecular cloud cores from which
they formed.

For each type of cloud (Figures \ref{acc_d0_w0} to \ref{acc_d1_w1}), 
we consider the evolution of `seed' binaries that form with 
initial mass ratios in the range $0.1 \leq q \leq 1.0$.  
We give the mass ratio $M_{\rm 2}/M_{\rm 1}$, separation $a$, ratio
of the mass of the circumbinary disc compared to the binary's mass
$M_{\rm cb}/M_{\rm b}$,
and relative accretion rate $\dot M_{\rm 2}/\dot M_{\rm 1}$,
as functions of the total mass that has fallen on to the binary 
system from the envelope, $M_{\rm acc}$.  
As mentioned in Section \ref{PBEcode},
the figures that are produced can be viewed in
two ways: either as giving the 
{\it evolution of individual protobinary systems} as
they evolve from their initial mass ($M_{\rm acc}=M_{\rm b}=1$) 
to a higher total mass
(up to $M_{\rm acc}=100$), or, for the first three panels in each figure, 
as giving the {\it final states} of binaries
that initially began with masses ranging from 1\% 
($M_{\rm acc}=100$ when the accretion stops) to 100\% 
($M_{\rm acc}=1$ if there is no accretion on to the protobinary) 
of the cloud's total mass $M_{\rm c}$.

\subsection{Solid-body rotation, uniform-density profile}
\label{soliddenuniform}

Figure \ref{acc_d0_w0} gives the results for binaries formed from initially
uniform-density clouds in solid-body rotation.  The long-term evolution of
the mass ratios is that they increase toward unity (equal mass components).
The separation initially decreases, due to accretion of gas with low mean 
specific angular momentum, but increases in the long-term. 
Both of these long-term effects are a consequence of the fact that 
the specific angular momentum of the infalling gas increases 
quickly as mass is accreted (Figure \ref{jrvsm}), 
since the accretion of material with high 
specific angular momentum increases the mass ratio and separation, while the
accretion of material with low specific angular momentum decreases both
the mass ratio and separation
(Artymowicz 1983; Bate 1997; Bate \& Bonnell 1997).

The different evolutions for the different `seed' mass ratios are in
large part due to the way we have chosen the initial conditions 
(Section \ref{PBEcode}).  
We assumed that the central region of the progenitor cloud from 
which the `seed' binary formed had the same angular 
momentum as that contained in the orbit of the 
`seed' binary (equation \ref{lbinary}).  For a `seed'
binary with a given mass ratio and separation these initial
conditions give the slowest possible rotation rate of the progenitor
core and, therefore, the {\it slowest possible evolution toward 
equal mass ratios and the formation of circumbinary discs}.  However,
they also mean that the gas which is accreted by a `seed'
binary with a low mass ratio has {\it less} specific angular momentum 
than for a binary with a high mass ratio.  Hence, the mass ratio of a `seed'
binary with a small mass ratio tends to decrease initially, while that of 
a binary with a large mass ratio tends to increase.

This dependence of the rotation rate of the progenitor cloud
on the mass ratio of the `seed' binary
also largely explains why the evolution of the separation 
(Figure \ref{acc_d0_w0}b) and circumbinary disc (Figure \ref{acc_d0_w0}c)
differ for different initial 
mass ratios.  In all cases, the separation decreases initially, but 
it does so quicker and for longer in systems with smaller initial mass 
ratios because the specific angular momentum of the infalling gas is lower.
Likewise, the formation of a circumbinary disc requires more accretion 
for systems with lower initial mass ratios.  However,
for the formation of a circumbinary disc, there is an additional effect.
For a system with a low mass ratio, the secondary is at a 
larger radius from the binary's centre of mass than in a system with a
high mass ratio.  Therefore, the infalling gas must have more
specific angular momentum before it can form a circumbinary disc 
rather than be captured by the secondary.

\begin{figure*}
\vspace{0.15truecm}
\centerline{\hspace{1.3truecm}\psfig{figure=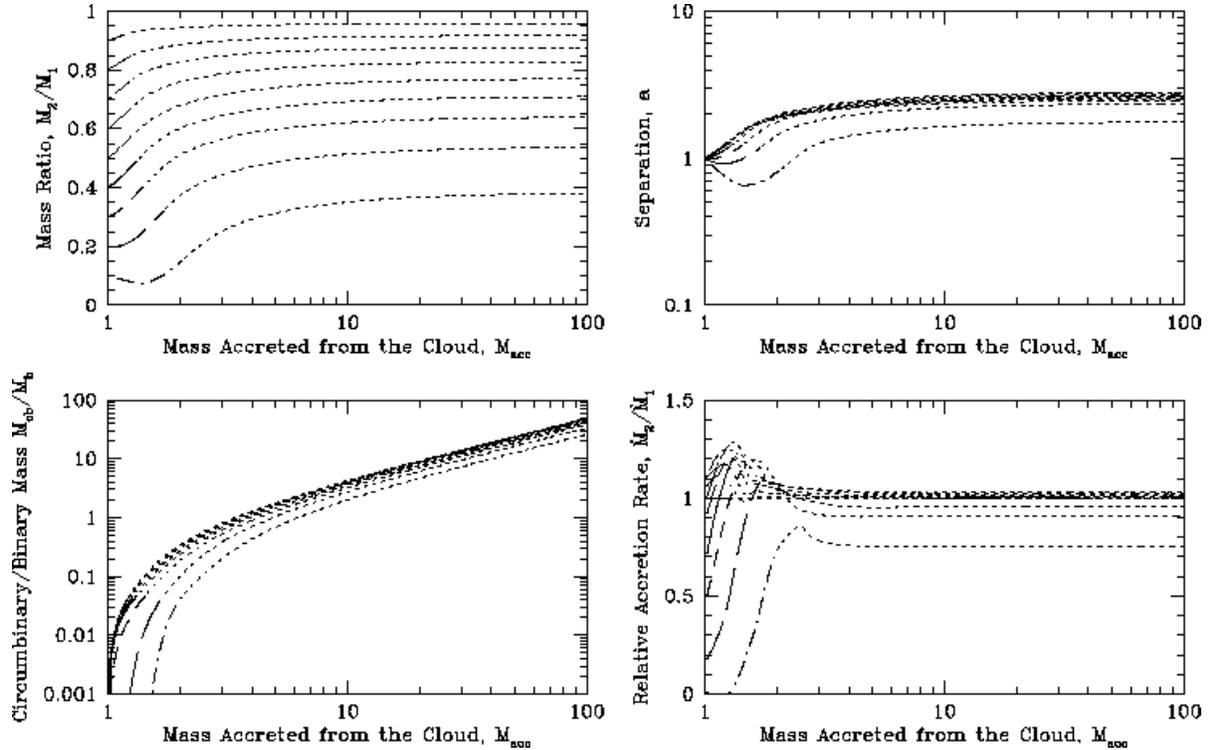,width=15.8truecm,height=9.9truecm,rwidth=17.0truecm,rheight=10.15truecm}}
\caption{\label{acc_d2_w0}  The evolution of protobinary systems which were formed in the centres of collapsing molecular cloud cores as they accrete from their gaseous envelopes.  The initial cores have $1/r^2$-density profiles and are in solid-body rotation.  The binaries rapidly evolve to a state where most of the infalling gas accumulates in a massive circumbinary disc and the binary itself ceases to evolve.  See Figure 9 for a description of the plots.  }
\end{figure*}

\begin{figure*}
\vspace{-0.7truecm}
\centerline{\hspace{0.2truecm}\psfig{figure=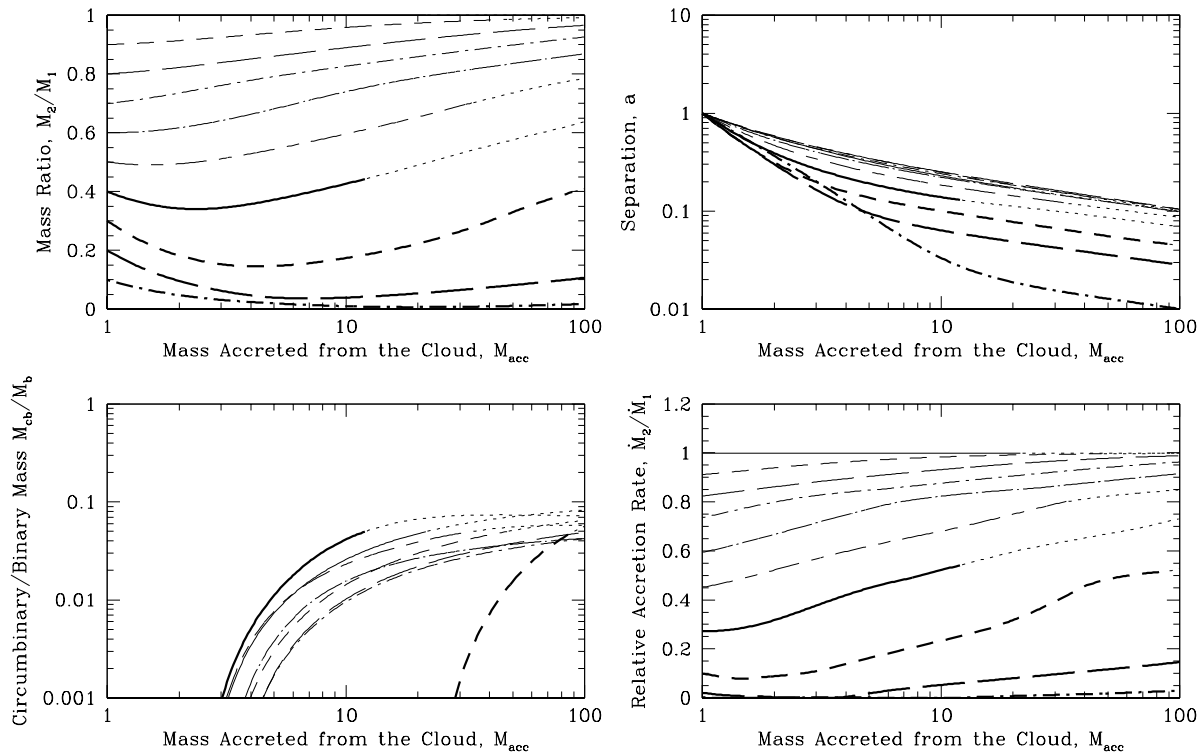,width=17.0truecm,height=17.0truecm,rwidth=17.0truecm,rheight=11.0truecm}}
\caption{\label{acc_d0_w1}  The evolution of protobinary systems which were formed in the centres of collapsing molecular cloud cores as they accrete from their gaseous envelopes.  The initial cores have uniform-density profiles and differential rotation of $\Omega_{\rm c} \propto 1/r$.  See Figure 9 for a description of the plots.  }
\end{figure*}

\begin{figure*}
\vspace{-0.7truecm}
\centerline{\hspace{0.2truecm}\psfig{figure=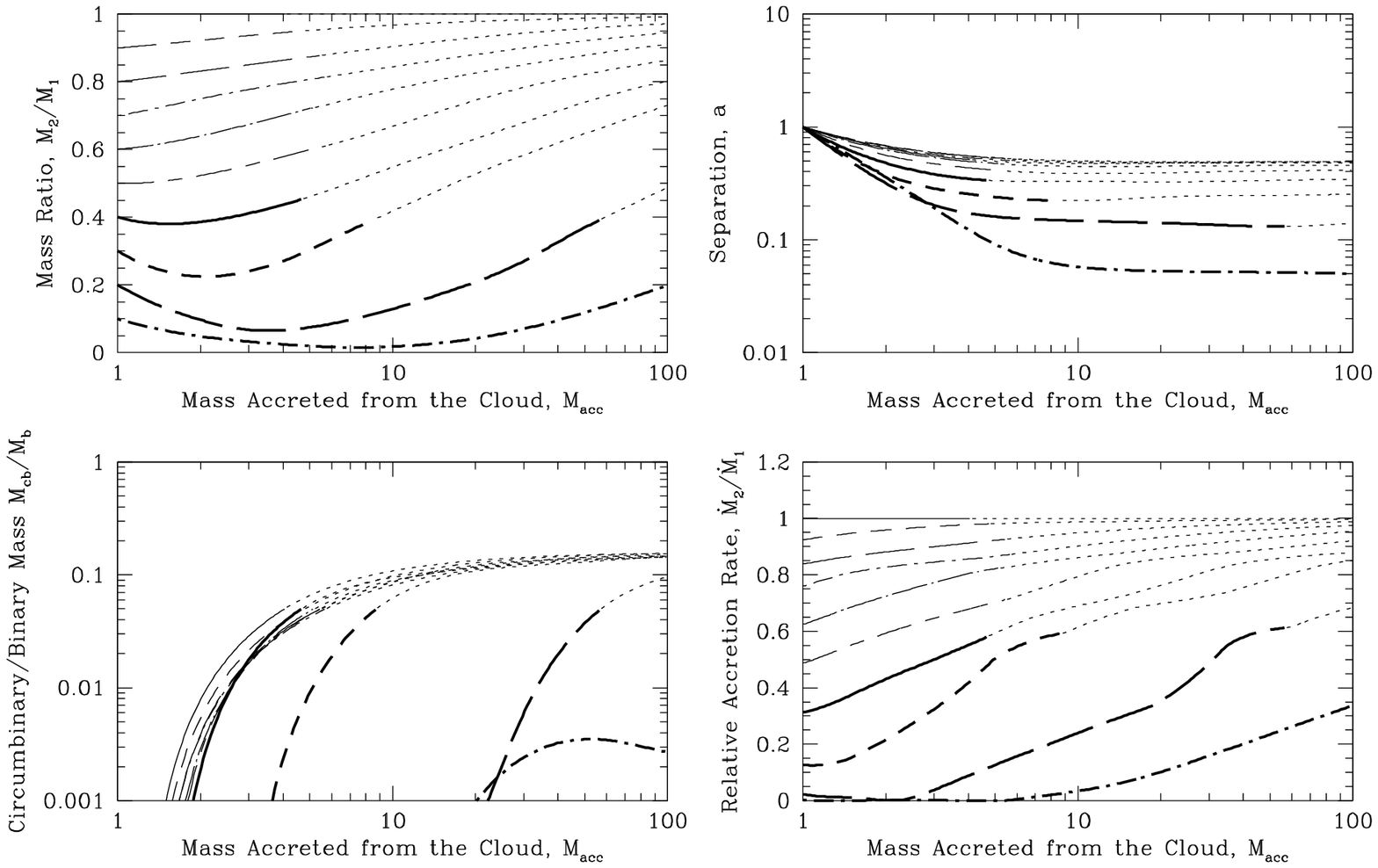,width=17.0truecm,height=17.0truecm,rwidth=17.0truecm,rheight=11.0truecm}}
\caption{\label{acc_d1_w1}  The evolution of protobinary systems which were formed in the centres of collapsing molecular cloud cores as they accrete from their gaseous envelopes.  The initial cores have $1/r$-density profiles and differential rotation of $\Omega_{\rm c} \propto 1/r$.  See Figure 9 for a description of the plots.  }
\end{figure*}

In all cases, the long-term evolution is that the binary reaches a
steady-state in which a constant fraction of the infalling gas 
is captured by the binary with the rest going into the circumbinary
disc and $\dot M_{\rm cb}/\dot M_{\rm b} \approx 0.17$.  Since the 
specific angular momentum of the gas that falls on to the binary is
always increasing (Figure \ref{jrvsm}), 
this indicates that a steady-state is established
in which the binary continually adjusts to the accretion so that ratio
of the specific angular momentum of the infalling gas to $\sqrt{GM_{\rm b}a}$
(which is related to the specific angular momentum of the binary) 
is constant (see equation \ref{jrel}).  Of course,
as shown by the test calculations (Section \ref{comparison}), 
when the circumbinary disc becomes
massive ($M_{\rm cb}/M_{\rm b} \simgreat 1/20$; dotted lines in Figures
\ref{acc_d0_w0} to \ref{acc_relax}) the 
interaction of the binary with the disc will not allow this same 
equilibrium to
be established.  Instead the separations will be lower 
and more of the infalling gas will settle into the circumbinary 
disc than is predicted here.  In addition, for the same increase 
in the mass of the binary, the mass ratio will tend to increase even 
more rapidly toward equal mass protostars because the smaller
separation of the binary means that the specific angular
momentum of the gas compared to that of the binary will be 
somewhat larger.  Thus, as with the assumption that the cloud rotates
at the slowest possible rate, {\it ignoring the interaction of the binary
with the circumbinary disc gives the slowest possible evolution 
toward equal masses and the formation of a massive circumbinary disc}.

\subsection{Solid-body rotation, 1/r-density profile}
\label{soliddenr}

Figure \ref{acc_d1_w0} gives the results for binaries formed from progenitor
clouds with $1/r$-density profiles in solid-body rotation.
The evolutions are similar to the uniform-density calculations, but
everything evolves toward high-angular-momentum behaviour after the
accretion of less gas.  
The mass ratios evolve toward unity and a massive circumbinary disc
is formed after less mass has been accreted.  
The separations begin increasing earlier, 
and increase by a greater amount for the same amount of accretion.

The more rapid evolution toward high-angular-momentum behaviour 
than with the uniform-density cloud is because the infalling gas 
has a more specific angular momentum initially, and its 
specific angular momentum increases more rapidly with mass
(Figure \ref{jrvsm}).  This also means that, to reach a 
steady-state, the binary must increase 
$\sqrt{GM_{\rm b}a}$ more rapidly.  Hence, a steady-state is 
attained when the infalling gas has more specific angular momentum 
relative to the binary than in the uniform-density case, and thus
$\dot M_{\rm cb}/\dot M_{\rm b}$ is also higher at $\approx 0.19$.
Again, however, the interaction of the binary with the circumbinary
disc, which is not taken into account by the PBE code, is likely to lead
to: significantly smaller separations; more mass in 
the circumbinary disc; and, for a given increase
in the binary's total mass, a mass ratio that is even closer to unity.

\subsection{Solid-body rotation, 1/r$^2$-density profile}
\label{soliddenr2}

Figure \ref{acc_d2_w0} gives the results for binaries formed from progenitor
clouds with $1/r^2$-density profiles in solid-body rotation.
The rate of increase of the specific angular momentum of the gas
as mass is accreted is now so rapid (Figure \ref{jrvsm}) 
that the binary cannot evolve fast enough to keep up with it.  
This results in the rapid formation and runaway growth of a 
massive circumbinary disc (Figure \ref{acc_d2_w0}c) because 
the infalling gas has too much angular momentum to be 
captured by the individual components of the binary.  
The binary only accretes a maximum of $1-3$ times its 
initial mass, even if the cloud mass is 100 times the 
initial binary mass -- the rest of the gas goes into the circumbinary disc.  

Of course, in reality, if the circumbinary disc becomes comparable 
in mass to the binary we would expect it to react in ways that are not 
possible to follow with the simple PBE calculations.  
Interaction between the binary and 
circumbinary disc could lead to two possibilities.
First, the disc may fragment to form one or more additional protostellar 
objects (e.g.~Bonnell \& Bate 1994a; Burkert \& Bodenheimer
1996; Burkert et al.~1997).  In this case, the model
has completely broken down, since in this paper we assume
that a binary is formed by an initial binary fragmentation and subsequent 
accretion.  There is no allowance made for additional
fragmentation.  The second possibility is that gas is accreted by
the circumstellar discs of the two protostars from the circumbinary 
disc (Artymowicz \& Lubow 1996; Lubow \& Artymowicz 1996).  If such 
accretion occurs, the binary's mass ratio will inevitably evolve 
rapidly toward unity because this material has high specific angular momentum
and will preferentially be captured by the secondary.  The mass ratios
would continue to follow the rapid increases seen early in 
Figure \ref{acc_d2_w0}a (when $M_{\rm acc} \simless 2$), giving even
faster evolution toward equal masses than was seen for 
cores with $1/r$-density profiles.

\subsection{Differential rotation, uniform-density profile}
\label{omegaruniform}

Figure \ref{acc_d0_w1} gives the results for initially 
uniform-density clouds in differential rotation with
$\Omega_{\rm c} \propto 1/r$.
The specific angular momentum of the infalling gas is 
low to begin with and only slowly increases as mass is accreted
(Figure \ref{jrvsm}).  Thus, a binary's mass ratio takes longer to
evolve toward unity, the separation continually decreases, 
and a circumbinary disc takes longer to form than in the solid-body case 
(Figure \ref{acc_d0_w0}).  Even after a binary has accreted 100 
times its initial mass, its mass ratio depends primarily on its
initial value (Figure \ref{acc_d0_w1}a).  The ratio of the mass in the 
circumbinary disc to the binary's mass is always less than 0.1, 
and in most cases less than 0.05, even after the infall of 100 times the
binary's initial mass.

\subsection{Differential rotation, 1/r-density profile}
\label{omegardenr}

Figure \ref{acc_d1_w1} gives the results for clouds with 
$1/r$-density profiles that are in differential
rotation with $\Omega_{\rm c} \propto 1/r$.  Again, with
differential rotation, the infall of gas with low specific
angular momentum is maintained for longer than with solid-body rotation.  
The mass ratios do all increase toward unity 
after more than 10 times the binary's initial mass has fallen in, however,
after 100 times the initial binary's mass has fallen in, 
a full range of mass ratios is still
possible.  The formation of a 
circumbinary disc is again delayed with differential rotation, 
although in most cases a steady-state is reached eventually with 
$\dot M_{\rm cb}/\dot M_{\rm b} \approx 0.15$.

\subsection{Differential rotation, 1/r$^2$-density profile}
\label{omegardenr2}

Finally, we consider the evolution of binaries formed from progenitor
clouds with $1/r^2$-density profiles that are in differential
rotation with $\Omega_{\rm c} \propto 1/r$.  From Figure \ref{jrvsm},
the relationship between angular momentum and mass in the progenitor
cloud is identical to a cloud with a $1/r$-density 
profile in solid-body rotation and, therefore, the evolution is
identical to that in Figure \ref{acc_d1_w0}.  

Compared to $1/r^2$-density with solid-body rotation, 
the effect of differential rotation is, once again, 
to decrease the rates of increase of the mass ratio and separation 
(at least while the binary is still accreting gas
in Figure \ref{acc_d2_w0}), and to delay the formation of a circumbinary
disc.  The reduced rate of increase of the specific angular momentum of the
infalling gas means that, unlike in the case
with solid-body rotation, the binary can reach an equilibrium with the
cloud; the runaway of the mass in the circumbinary disc is avoided.
However, even such strong differential rotation is not enough
to stop the mass ratios being driven toward unity with an initial
density profile of $\rho \propto 1/r^2$.

\section{Relaxing some of the assumptions}
\label{relax}

\begin{figure*}
\vspace{-0.7truecm}
\centerline{\hspace{0.2truecm}\psfig{figure=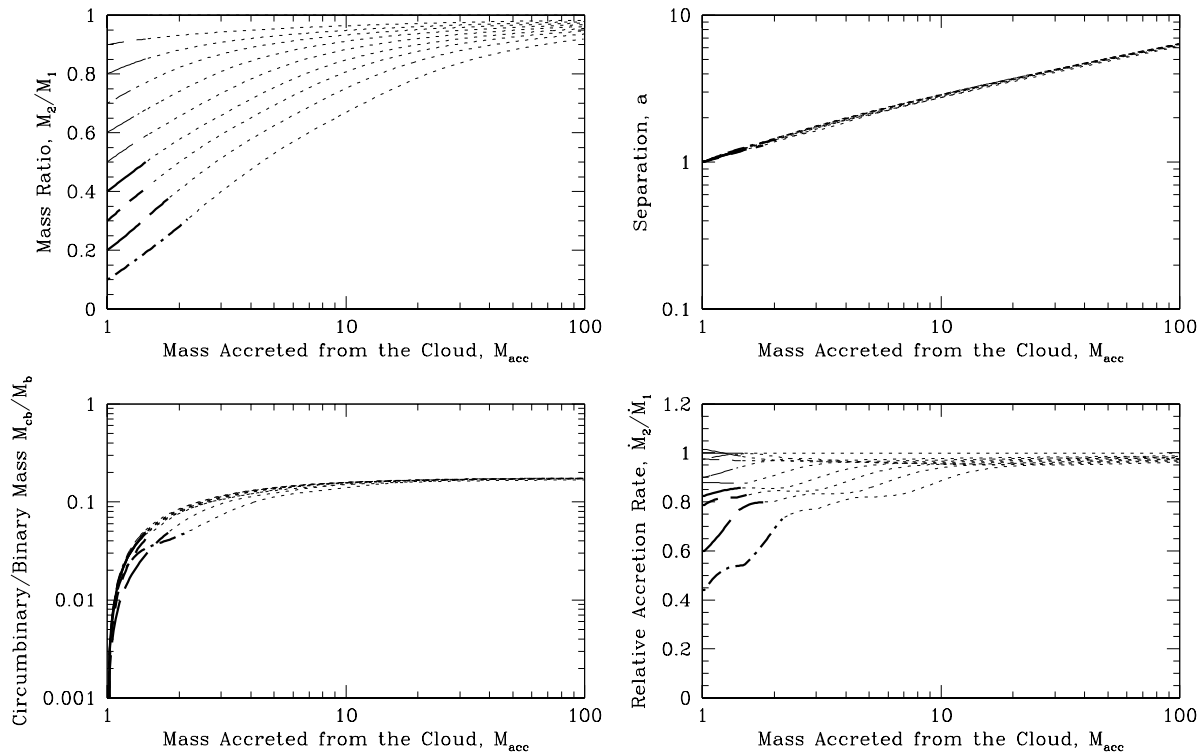,width=17.0truecm,height=17.0truecm,rwidth=17.0truecm,rheight=11.0truecm}}
\caption{\label{acc_relax}  The evolution of protobinary systems which were formed in the centres of collapsing molecular cloud cores as they accrete from their gaseous envelopes.  As in Figure 9, the initial cores have uniform density profiles and are in solid-body rotation, but here the clouds are rotating more rapidly and are all rotating at the same rate, regardless of the binary's initial mass ratio (see Section 5.1).  This type of evolution is appropriate for binaries that form via the fragmentation of a massive circumstellar disc surrounding single protostar.  See Figure 9 for a description of the plots.  }
\end{figure*}

\subsection{The rotation rate of the progenitor cloud}
\label{rotationrate}

In the previous section, we assumed that the orbital angular momentum
of the `seed' binary system, $L_{\rm b}$, was equal to the 
angular momentum of the spherical central region of the cloud, $L_{\rm cen}$, 
from which the `seed' binary formed (equation \ref{lbinary}).  
This assumption gives a {\it lower limit} to the rate of rotation of the
progenitor cloud, since it does not allow for the possibility 
of circumstellar discs around the `seed' binary which would
contain additional angular momentum.  Furthermore, if the 
`seed' binary was formed via fragmentation of a circumstellar 
disc surrounding a single 
protostar, then the disc must have had a radius at least as big as the
separation of the `seed' binary which forms (i.e.~at least 
some gas in the disc must have had a specific angular momentum of 
$j=\sqrt{GM_{\rm b}a}$).  Thus, some of the gas which falls on to 
the system from the envelope immediately after the `seed'
binary forms is expected to have at least this much specific
angular momentum.  As a comparison, the first gas to be accreted by the
binaries in Figure \ref{acc_d0_w0} had at most $0.21-0.63$ of this 
value, depending on the initial mass ratios.  
Therefore, it is conceivable that the progenitor clouds may be 
rotating faster than we assumed in the previous section.

To demonstrate the effect of the progenitor cloud having a greater
rate of rotation, we performed calculations which evolved binaries whose
progenitor clouds were rotating rapidly enough that some (that part of
the envelope which is in the plane of the binary) of the first gas to
fall on to the binary had a specific angular
momentum of $j=\sqrt{GM_{\rm b}a}$.  Thus, rather than setting 
$R_{\rm b}^2 \Omega_{\rm Rb}$ using equation \ref{lbinary}, we set
\be
R_{\rm b}^2 \Omega_{\rm Rb} = \sqrt{G M_{\rm b} a},
\ee
{\it regardless of the initial mass ratio of the binary}.
Thus, all the progenitor clouds have the same initial 
density profile and rotation rate and all the `seed' binaries have
the same separation, regardless of the mass ratio of the `seed' binary.  
This also allows us to examine the degree to which the evolution 
of a binary depends on its mass ratio, and how much it depends on 
the properties of the gas which it accretes from the infalling envelope.
These `seed' binaries could have formed via the fragmentation 
of a disc, where we assume that the angular momentum that 
is in excess of the `seed' binary's orbital angular 
momentum, $L_{\rm b}$, is contained in circumstellar discs.

The results for an initially uniform-density cloud in solid-body
rotation are given in Figure \ref{acc_relax}.  Due to the greater 
rotation rate of the progenitor cloud and, hence, the higher specific 
angular momentum of the gas which falls on to the binary, 
the mass ratios are driven toward unity after much less mass has been
accreted compared with when the slowest possible
rotation rate was used (Figure \ref{acc_d0_w0}).
Thus, {\bf the rates of increase toward mass ratios
of unity that are given in Section \ref{evolution} are lower limits}.

In contrast to the previous section, the evolutions of the 
separation and the mass in a circumbinary disc are quite independent 
of the mass ratio of the binary, demonstrating that {\bf the evolution
of a binary's separation and the formation of a circumbinary disc 
depend almost entirely on the properties of the infalling gas;
the mass ratio of the binary has almost no effect.}  
The very small dependence of the mass of the circumbinary disc 
on the binary's mass ratio
is because the secondary is farther from the centre of mass 
in a binary with a lower mass ratio.

\subsection{Different density profiles}

The evolutions presented in Section \ref{evolution} are for 
idealised progenitor cloud cores with power-law density and 
rotation profiles.  These clearly illustrate the main dependencies of 
the evolution of a protobinary system as it accretes to its final
mass.  However, the PBE code easily be applied to other types 
of progenitor cloud core.
For example, we have performed evolutions beginning with Gaussian
centrally-condensed cores with inner to outer density contrasts of 20:1.
These have density profiles that are similar to observed pre-stellar 
molecular cloud cores in that they are steep on the outside and flatter
near the centre (Ward-Thompson et al.~1994;
Andr\'e, Ward-Thompson, Motte 1996; Ward-Thompson, Motte, \& Andr\'e 1999).
As one might expect, the evolutions lie between those for uniform-density
clouds and those with $\rho \propto 1/r$, although they are 
closest to the uniform-density case.

\subsection{Eccentric binary systems}
\label{eccentricity}

In this paper, strictly, we only consider the evolution 
of protobinary systems with circular orbits.  However, in the test cases
of Section \ref{comparison}, where we compared the evolution given by 
the PBE code with those from full SPH calculations, we obtained
good agreement even though the orbit of the binary in the SPH 
calculation had eccentricities up to $e \approx 0.2$.  Therefore, we are
confident that the results in this paper are valid for binaries with 
eccentricities $e \simless 0.2$.

For larger eccentricities, we expect the evolution of the binaries
should be {\bf qualitatively similar} to what we have found for circular
binaries.  Hence, we still expect that the long-term effect of accretion
will be to drive the mass ratios toward unity and form circumbinary
discs and that this evolution will be enhanced with more 
centrally-condensed initial conditions and diminished by differential
rotation.

However, there will be quantitative differences.  Most importantly,
the formation of a circumbinary disc will be delayed 
in an eccentric system for two reasons.  First, for the same 
semi-major axis, an eccentric binary has less angular momentum than a
circular one and, thus, the gas from which it first formed (and hence
the molecular cloud core as a whole) may have been rotating more slowly.  
This is would also mean that more gas is able to be accreted 
before the binary's mass ratio is driven toward unity.
Second, for binaries with the same semi-major axis separation, 
a circumbinary disc must be formed at a larger radius from the 
centre of mass for an eccentric binary than for a circular binary 
to avoid disruption (e.g.~Artymowicz \& Lubow 1994).  
Thus, eccentric systems are expected to be able to accrete significantly more 
material than circular binaries before forming a circumbinary disc. 

Other effects may include collisions between the protostars and their
circumstellar discs (e.g.~Hall, Clarke \& Pringle 1996) 
and, of course, the eccentricity itself 
is expected to evolve due to accretion and the interactions 
between the protostars and the 
discs.  These effects are beyond the scope of this paper, but 
would certainly be well worth studying.

\section{The Properties of Binary Systems}
\label{predictions}

The aim of this paper is to determine, for a particular 
model of binary star formation, the properties of binaries and
how they depend on the initial conditions in their progenitor 
molecular cloud cores.  By comparing these predictions with 
observations, we hope to determine whether this is a viable 
model for binary star formation and, if so, to constrain the 
initial conditions for binary star formation.
In order to do so, however, we must first discuss the 
relationship that is expected between the mass of a `seed' 
binary system and its separation.

\subsection{Dependence of the mass of a `seed' protobinary on its separation}
\label{bmass_vs_sepsec}

In this paper, we consider the evolution of a `seed' binary system 
as it accretes gas from an infalling envelope, but we do not specify 
exactly how the protobinary is formed.  Presumably, it is formed
via some sort of fragmentation process (see Section 1).
In order for binary fragmentation to occur, the Jeans radius at the 
time of fragmentation must be less than or approximately equal to the 
separation of the protobinary that is formed.  Thus, 
$a \simgreat 2 R_{\rm J}$, where $R_{\rm J}$ is the Jeans radius
\be
\label{jeansradius}
R_{\rm J} \approx \frac{2 G M_{\rm frag} \mu}{5 R_{\rm g}T},
\ee
$\mu$ is the mean molecular weight, and $R_{\rm g}$ is the gas constant.
The `seed' binary's initial mass is approximately twice the mass
of each fragment, $M_{\rm frag}$.  Therefore, for a constant
temperature, $T$, we expect a roughly linear 
relationship between the mass of a `seed' binary and its separation:
$M_{\rm b} \approx 2 M_{\rm frag} \propto a$.

\begin{figure}
\vspace{-0.2truecm}
\centerline{\hspace{0.2truecm}\psfig{figure=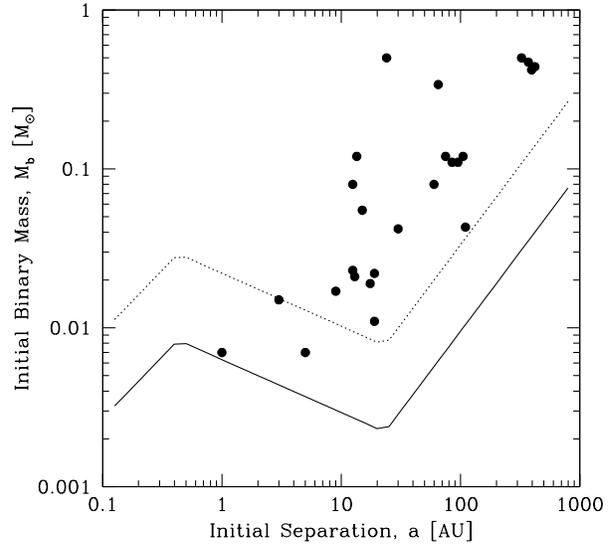,width=8.0truecm,height=8.0truecm,rwidth=8.0truecm,rheight=7.5truecm}}
\caption{\label{bmass_vs_sep}  The dependence of a protobinary's initial mass on its separation.  The points give the results from 26 fragmentation calculations (Figure 13, Boss 1986).  The solid and dotted lines give simple estimates (see Section 6.1) of the minimum mass that a `seed' protobinary system should have as a function of its separation.}
\end{figure}

Figure \ref{bmass_vs_sep} gives the initial mass of a binary versus 
its separation from a series of fragmentation calculations performed by
Boss \shortcite{Boss86}.  Indeed, there is a strong, approximately linear, 
relationship between the binary's mass and its separation.

Using equation \ref{jeansradius}, we plot (solid line) the 
relationship if $a = 2 R_{\rm J}$.  We use $\mu=2.46$ with
$T=10$ K for densities below $10^{-13}$ g cm$^{-3}$ and 
$T \propto \rho^{\gamma-1}$ with $\gamma=1.4$ for 
$10^{-13} \leq \rho < 6\times 10^{-8}$ g cm$^{-3}$ and 
$\gamma=1.1$ for $\rho \geq 6\times 10^{-8}$  g cm$^{-3}$ (e.g.~Tohline
1982).  The numerical results generally lie above the solid line
because the properties of the binaries were calculated somewhat after they
first began to form and they have already accreted some gas.
For example, if we assume that the fragments are embedded in a sphere of 
gas with radius $2 R_{\rm J}$ and a mean density of half the 
mean density of the fragments which is rapidly accreted by the
fragments, then we obtain the dotted line for the masses of 
the `seed' binaries.
Furthermore, the fragments typically form on elliptical orbits 
and are initially falling toward each other (which moves
the numerical results to smaller separations).  

Generally, for separations $\simgreat 10$ AU, wider `seed' 
binaries have larger initial masses, while for separations
$\simless 10$ AU the initial masses are $\approx 0.01 {\rm M}_\odot$.
Therefore, for example, in order to form binary systems with a solar-mass 
primary, close systems ($\simless 10$ AU) may have to 
accrete $\sim 100$ times their initial mass.  Systems with separations 
$10 \simless a \simless 100$ AU need to accrete between $100-10$ times
their initial mass, and the widest systems need only accrete 
a few times their initial mass.

This leads us to our first prediction: for binary systems with the 
same total mass, {\bf
the properties of close systems will be more heavily
determined by accretion than those of wide systems}.  In turn this means
that the mass ratios of wide binary systems are expected to be 
determined primarily from the initial density structure in the 
molecular cloud cores and so they may enable us to better understand the
initial conditions for star formation by giving us a way to measure
the density structure.  Therefore,  
{\bf in order to study the initial conditions
for star formation we should consider the properties of wide binaries,
while to learn more about the evolution due to accretion it is best to
consider close binaries}.

A further prediction is: {\bf if high and low mass stars form by the
same general process, and the initial mass of a `seed' binary
is independent of the mass of the progenitor core, then the properties
of massive binary systems should be more heavily influenced 
by accretion than those of lower mass systems with the same separation}.

\subsection{Binary mass ratios}

In sections \ref{evolution} and \ref{relax}, we found that
{\bf the long-term evolution of an accreting binary is for its 
mass ratio to approach unity}.

\subsubsection{Dependence of the mass ratio on separation}

As argued in Section \ref{bmass_vs_sepsec}, for 
molecular cloud cores of a given mass, the amount of mass accreted 
by a binary, relative to its initial fragmentation mass, will be greater
for close binaries than for wide systems.  Thus, {\bf closer binary
systems are more likely to have mass ratios near unity than wider 
systems with the same total mass}.

Duquennoy \& Mayor \shortcite{DuqMay91} surveyed main-sequence 
G-dwarf stellar systems.  They found that the mass ratio 
distribution, averaged over binaries with all separations, increases toward
small mass ratios.  However, there is mounting evidence that
the mass-ratio distributions differ between short and long-period
systems with the distribution for close binaries ($P <
3000$ days; $a \simless 5$ AU) consistent with
a uniform distribution (Mazeh et al.~1992; Halbwachs, Mayor \& Udry~1998).  
Thus, relative to wide systems, the close systems are biased toward
mass ratios of unity, in agreement with the above prediction.

The fraction by which the mass of a `seed' binary must
be increased in order for its mass ratio to approach unity depends
on the conditions in the progenitor cloud core.  In general, {\bf the
less centrally-condensed a core is and/or the more differential rotation
it has ($\beta \leq 0$), 
the easier it is to form binary systems with low mass ratios
for a given increase in the binary's initial mass}.  By inquiring how
easy it is to reproduce the observed mass-ratio distributions with
different types of progenitor molecular cloud core, 
we can use these results to constrain the initial conditions for binary
star formation.

Duquennoy \& Mayor \shortcite{DuqMay91} found that binaries containing
G-dwarfs with
separations $\simgreat 30$ AU generally have unequal mass ratios 
(typically $q\approx 0.3$).  From Figure \ref{bmass_vs_sep}, typically,
these binaries may be expected to accrete $\sim 10$ times their 
`seed' mass.  For uniform-density progenitor 
clouds in solid-body rotation
(Figure \ref{acc_d0_w0}a), this observed mass 
ratio distribution could be produced if the fragmentation process 
typically produces `seed' binaries with low mass ratios ($q \approx 0.2$).
However, with more centrally-condensed cores it becomes progressively
more difficult to produce the observed mass-ratio distribution because
the accretion drives the initial mass ratios toward unity
(Figures \ref{acc_d1_w0}a and \ref{acc_d2_w0}a).  With
$\rho \propto 1/r$ cores, the `seed' binaries must typically have
mass ratios of $q \approx 0.1$ in order to produce G-dwarf binaries
that have typical mass ratios of $q\approx 0.3$.  With 
$\rho \propto 1/r^2$ (assuming
the `seed' binaries manage to accrete from their circumbinary discs) it
is difficult to see how the observed mass-ratio distribution could be
produced because of the rapid evolution toward equal mass protostars.  
If the progenitor cores have significant differential
rotation, uniform-density and $\rho \propto 1/r$ cores can easily
produce the observed mass ratios, but it is still difficult to 
produce G-dwarf binaries with low mass ratios from cores with 
$\rho \propto 1/r^2$.

For close G-dwarf binary systems ($a \simgreat 5$ AU), the constraints are
even more pronounced.  Typically, we expect such binaries to accrete
$\approx 100$ times their `seed' mass, yet the observed 
mass-ratio distribution is approximately flat (i.e.~approximately
1/2 the binaries have $q<0.5$).  It is effectively impossible
for cores in solid-body rotation to produce such a mass ratio 
distribution if they are significantly centrally-condensed.  Even for
uniform-density cores, approximately 1/2 of the `seed' binaries
would need to have $q<0.1$, although taking into account the effect of
eccentricity (Section \ref{eccentricity}) it is probable that cores
with nearly uniform-density profiles can reproduce the observations.
If the cores have significant differential rotation, the observed
mass ratios can be produced with cores that are as 
centrally-condensed as $\rho \propto 1/r$, but cores with
$\rho \propto 1/r^2$ are still excluded.

We note that, as discussed in Section \ref{relax}, the results 
in Section \ref{evolution} give the slowest possible evolution 
toward mass ratios of unity (unless the binary has a significant
eccentricity).  Thus, it may be even more difficult to form 
unequal mass binaries than suggested by the above discussion.
Even taking the above numbers, with solid-body rotation the mass
ratios of the `seed' binaries must typically be $q \approx 0.1-0.2$.
To obtain such mass ratios via direct fragmentation requires
that the progenitor molecular cloud cores have significant 
asymmetries (e.g.~Bonnell \& Bastien 1992) which implies that they
are formed and/or triggered to collapse by dynamical processes.
Low mass ratios can also be obtained via rotational fragmentation
(Section \ref{introduction}), but then the binary's mass ratio
will be driven toward unity even more rapidly.  Thus, in reality,
molecular cloud cores may have a degree of differential rotation.

In summary, {\bf the observed mass-ratio distributions are most
easily explained by cores that have density profiles that are
less centrally condensed than $\rho \propto 1/r$ (e.g.~Gaussian),
possibly with a small amount of differential rotation.}  
This is in good agreement with the observed density profiles of
pre-stellar cores 
(Ward-Thompson et al.~1994; Andr\'e, Ward-Thompson, Motte 1996; 
Ward-Thompson, Motte, \& Andr\'e 1999).  If 
the progenitor cores rotate as solid-bodies it is essentially 
impossible to produce
the observed mass-ratio distribution of close binaries if the
cores are much more centrally condensed than a Gaussian-density 
profile.  If differential rotation is allowed, cores with density
profiles as steep as $\rho \propto 1/r$ are feasible (with 
strong differential rotation), but cores with
$\rho \propto 1/r^2$ still cannot reproduce the observations.

\subsubsection{Dependence of the mass ratio on the binary's total mass}

If all binary systems form by the same general process, regardless 
of a system's final total mass (i.e.~from the collapse of molecular
cloud cores with the same properties, only more massive), and if
the initial mass of a `seed' binary (Section \ref{bmass_vs_sepsec})
is independent of the mass of the core (depending only
on its initial separation), then {\bf massive binaries should have
more-equal mass ratios than low-mass binaries of the same separation}.
This is because systems with a larger final mass would have had to
accrete more, relative to their initial mass, than systems with a lower
final mass.

Unfortunately, unbiased surveys of stars more massive than G-dwarfs
are only slowly becoming available.  Preliminary results for O-star 
binaries (e.g.~Manson et al.~1998) show a difference in the 
mass-ratio distribution between close 
($P \simless 40$ days; $a \simless 1/2$ AU) 
and wide ($P \simgreat 10^4$ years; $a \simgreat 1000$ AU) 
systems, with close systems biased toward mass ratios of 
unity, but comparison between
the mass-ratio distributions of high and low-mass binaries is not
yet possible.  In addition, Bonnell, Bate \& Zinnecker (1998) recently
proposed that O-stars form in a different way to lower mass stars,
from the collision of less massive stars in very dense star-forming 
regions.  This theory predicts a large frequency of close binaries for
O-stars (due to tidal capture) with a mass-ratio distribution that
is not determined by accretion from an infalling envelope.  Thus, we
strongly encourage surveys that will determine the mass-ratio distributions
of B-, A-, and F-stars.

\subsubsection{Formation of brown dwarf companions to solar-type stars}
\label{browndwarfs}

Given that it becomes more difficult to form binaries with low mass ratios
as more gas is accreted by a binary, 
it is of interest to ask how easy it is to form stars with brown-dwarf 
companions (see also Bate 1998).  Since wider binaries should 
have to accrete less gas, relative to their initial mass, to 
attain the same final total mass, then {\bf the frequency of 
brown dwarf companions to solar-type stars should increase with separation}.
Likewise, we expect that systems with lower final masses also accrete
less relative to their initial masses so that {\bf for the same range of
separations, brown dwarf companions should be more frequent in 
systems with a lower total mass than in higher-mass systems.}

Taking the types of cores that most easily reproduce 
the mass-ratio distributions
of low-mass stars (i.e.~near-uniform density cores in solid-body rotation),
we find that a brown dwarf companion to a solar-type star could be formed 
if an extreme mass ratio ($q \approx 0.1$) was produced at fragmentation and
the system subsequently increased its mass by a factor of $\sim 20$ or
less (Figure \ref{acc_d0_w0}a).  This implies that it may be quite easy to form
brown dwarf companions to stars of around a solar mass or less 
with separations $\simgreat 10$ AU.  Indeed, two wide systems have 
been found: GL 229B
is a brown dwarf companion ($0.02 - 0.06~ {\rm M}_\odot$) to a $0.6~
{\rm M}_\odot$ M-dwarf with a separation of $\approx 50$ AU (Nakajima
et al.~1995); G 196-3B is a brown dwarf companion ($0.02 - 0.04~ 
{\rm M}_\odot$) to a $0.4~{\rm M}_\odot$ 
M-dwarf with a separation of $\approx 300$ AU (Rebolo et al.~1998).

However, close systems ($\simless 10$ AU) need to
accrete $\approx 100$ times their initial mass in order to obtain a
solar-mass primary and, due to the evolution of the mass ratio toward
unity, it would be very difficult for the secondary to have the 
final mass of a brown dwarf after this amount of accretion.
(Figure \ref{acc_d0_w0}a).  Therefore, {\bf brown-dwarf companions
to stars with masses $\approx 1 {\rm M}_\odot$ and separations
$\simless 10$ AU are likely to be extremely rare or perhaps even nonexistent}.
This prediction is supported by the recent radial-velocity searches 
for giant planets (see Marcy, Cochran \& Mayor 1999 and references within).
Although many planetary candidates 
($M \sin i \simless 0.013~{\rm M}_\odot $) have been found with 
separations less than a few AU, there are currently only 
4 brown dwarf ($0.013~ {\rm M}_\odot \simless 
M \sin i \simless 0.075~ {\rm M}_\odot $) candidates from $\approx 600$
target stars and even these could be stellar companions with orbits nearly
perpendicular to the line of sight
(Marcy, Cochran \& Mayor 1999).

Within this model, the easiest way to obtain such close systems with extreme 
mass ratios would be that they formed from cores with significant 
differential rotation, or that the companion was formed significantly 
{\it after} the primary.  In the latter case, the primary would already 
have a significant fraction of its final mass and, hence, the protobinary 
would not have to increase its total mass by such a large factor.
However, if this was achieved via the 
fragmentation of a circumstellar disc, then the fragmentation
would have to occur after the primary and its disc had accreted 
a large fraction of the envelope, otherwise the secondary
would still end up with a stellar mass due to subsequent accretion 
(Section \ref{rotationrate} and Figure \ref{acc_relax}a).

\subsection{Binary separations}

In Section \ref{evolution}, we found that a binary's separation can
evolve significantly due to accretion, increasing 
or decreasing by up to 2 orders of magnitude (Figures \ref{acc_d0_w0}b to
\ref{acc_relax}b).  However, for most cases, if the binary's 
separation is increasing the PBE code also predicts that a 
massive circumbinary disc will be present.  In test case 2
(Section \ref{testcase2}; Figure \ref{PBESPH2}), we found that if a
massive circumbinary disc is present the interaction of the binary
with the circumbinary disc is likely to negate the separation-increasing
affect of the accretion and the separation will remain approximately 
constant.

Therefore, {\bf we conclude that a binary's separation is likely to 
decrease or remain of the same order as 
its initial value during the accretion of
the gaseous envelope}.  After the accretion phase, if a binary has a
circumbinary disc its separation is expected to decrease.  Without a
circumbinary disc, its separation is likely to increase as the circumstellar
discs evolve viscously and transfer their angular momentum to the orbit
of the binary.  However, the angular momentum contained in the 
circumstellar discs is likely to be small compared to the orbital angular
momentum of the binary and, therefore, the binary's separation is expected to
increase by less than a factor of two.

\subsection{Circumstellar discs}

Bate \& Bonnell \shortcite{BatBon97} studied the disc formation 
process in accreting protobinary systems and established criteria 
for the formation of circumstellar and circumbinary discs.  
They found that if a protobinary system only accretes 
gas with low specific angular momentum after its formation, 
the primary will have a circumstellar disc but the secondary may not.
The reverse is not true; if a circumstellar disc is formed
around the secondary, the primary will also have a disc.  
These conclusions can also be seen in the snapshots from the SPH test cases in 
Figures \ref{UD_S2}, \ref{UD_B1}, \ref{1r_B1}, and \ref{1r_D1}.

With the PBE code, we do not differentiate between gas that is directly
accreted by a protostar and gas that is captured in its circumstellar
disc (presumably to be accreted by the protostar at a later time).  
However, we do
determine the relative accretion rate on to the secondary and 
its circumstellar disc compared to the primary and its 
circumstellar disc $\dot M_2/\dot M_1$ during formation of a binary
(Figures \ref{acc_d0_w0}d to \ref{acc_relax}d).  We find that the
relative accretion rate is $\dot M_2/\dot M_1 \leq 1.3$
and in the majority of cases is less than unity.  Therefore,
{\bf we expect that in most cases the circumsecondary disc will have 
a mass that is less than or similar to that of the circumprimary disc.}  
This conclusion is valid
unless the circumprimary disc accretes on a shorter time-scale than
the circumstellar disc.  In fact, Armitage, Clarke \& Tout (1999)
showed that the circumsecondary disc is expected to accrete {\it more rapidly}
than the circumprimary disc and, therefore, this effect should only
be enhanced.  This conclusion is in excellent agreement with the latest 
observations of circumstellar material around young binary systems.
Ghez, White \& Simon \shortcite{GWS97} considered UV and NIR 
excess emission from the components of young binary systems 
and found that the excesses are either comparable or 
dominated by the primary, suggesting that the gas in 
circumstellar discs is either distributed similarly or 
preferentially around the primary.

The only cases where a circumsecondary disc may become significantly
more massive than the circumprimary disc are those where the 
gaseous envelope has effectively been exhausted and accretion 
on to the circumstellar discs comes primarily from a circumbinary disc.  
In these cases, because the 
gas has very high angular momentum with respect to the binary, it will
preferentially be captured by the secondary (Artymowicz \& Lubow 1996; 
Bate \& Bonnell 1997).

\subsection{Circumbinary discs}
\label{circumbinary}

From the results in Section \ref{evolution}, just as the mass 
ratio of an accreting protobinary evolves toward unity in the long-term, 
the more material that is accreted by a protobinary, the more 
likely it is that a circumbinary disc is formed.
Therefore, following the same argument that we made for the mass ratios
of binary stars, {\bf we predict that 
closer binary systems are more likely to have circumbinary
discs than wider systems with the same total mass}.  
Furthermore, if massive and low-mass binary
systems via the same process, and if the initial mass of
a `seed' binary is independent of the mass of the core, 
then  {\bf massive binaries are more likely to have 
circumbinary discs than low-mass binaries of the same separation}.

The first of these predictions is in good agreement with recent observations.
Jensen, Mathieu \& Fuller \shortcite{JMF96} surveyed 85 
pre-main-sequence binary systems and found that, while emission 
presumably associated with circumbinary discs could be found 
around many close binaries (separations less than a few AU), 
circumbinary emission around binaries with separations of a
few AU to $\approx 100$ AU is almost entirely absent.  The only
exception was GG-Tau with a separation of $\approx 40$ AU \cite{DutGuiSim94}.
Furthermore, Dutrey et al.~(1996) performed an imaging survey of 18 
multiple systems that could resolve circumbinary discs with radii
$\simgreat 100$ AU and found only one circumbinary disc around UY-Aur
which has a separation of $\approx 120$ AU.

As with the mass-ratio evolution, the quantitative predictions depend on
the properties of the progenitor cloud cores: 
{\bf the more centrally-condensed a core is and/or the
less differential rotation it has ($\beta \leq 0$), the lower the 
fraction by which a binary has to increase its mass before a circumbinary
disc is formed}.

For cores in solid-body rotation, a uniform-density cloud leads 
to a significant circumbinary disc ($M_{\rm cb}/M_{\rm b} > 0.05$) 
after a circular binary accretes $\approx 2-40$ times its initial mass 
(depending on the initial mass ratio; Figure \ref{acc_d0_w0}c).  Using
Figure \ref{bmass_vs_sep}, we would expect most {\it circular} binaries with 
separations $a \simless 100$ AU to have circumbinary discs, while many
with separations $a \simgreat 100$ AU should not have circumbinary discs.
However, most wide binaries have large eccentricities
(e.g.~Duquennoy \& Mayor 1991) and more material must be accreted
by an eccentric binary before a circumbinary disc is formed 
(Section \ref{eccentricity}).  Taking this into account, uniform-density
cores in solid-body rotation are likely to produce 
circumbinary discs around most binaries with $a \simless 10$ AU
and some binaries with intermediate separations.  However,
very few should exist around binaries with separations $a \simgreat 100$ AU,
in reasonable agreement with the observations.

For a core with $\rho \propto 1/r$ (Figure \ref{acc_d1_w0}), 
a binary can only increase its mass by a factor of $\approx 1.5-6$
before a circumbinary disc is formed.  In this case, most 
binaries with separations $a \simless 100$ AU would be expected to 
have circumbinary discs, even taking eccentricity into account.
With a $1/r^2$-density profile a binary can't even double its mass before
a circumbinary disc is formed (Figure \ref{acc_d2_w0}) so that almost
all binaries would be expected to have circumbinary discs, in strong 
disagreement with observations.

Differential rotation allows a binary to accrete more mass before 
a circumbinary disc is formed.  In most cases, binaries formed from 
uniform-density cores (Figure \ref{acc_d0_w1}) can accrete up to 100 times 
their initial mass without forming a massive circumbinary disc, 
meaning that even the closest binaries would not have circumbinary discs.  
Cores with $1/r$-density profiles (Figure \ref{acc_d1_w1}) allow 
from $5-100$ times a binary's initial mass to be accreted so that
most binaries with separations $\simless 10$ AU and many with separations 
$\simless 100$ AU would have circumbinary discs.  Cores with 
$1/r^2$-density profiles would still produce discs around most
binaries with separations $a \simless 100$ AU.

Therefore, as with our predictions concerning binary mass-ratio
distributions, {\bf the circumbinary disc observations can be 
reasonably well explained
if most binaries form from progenitor cores which are less centrally
condensed than $\rho \propto 1/r$ (e.g.~Gaussian), possibly with a
small amount of differential rotation}.
Cores with $\rho \propto 1/r$ are possible if there is significant
differential rotation, but the singular isothermal sphere ($\rho \propto
1/r^2$) cannot reproduce the observations even with extreme differential
rotation.

\section{Conclusions}
\label{conclusions}

We have considered a model for the formation of binary stellar
systems which has been inspired by the results obtained from 
$\approx 20$ years of
study of the fragmentation collapsing molecular cloud cores.
In the model, a `seed' protobinary system forms, presumably via
fragmentation, within a collapsing molecular cloud core and evolves
to its final mass by accreting material from an infalling 
gaseous envelope.  

We developed and tested a method which can rapidly follow the
evolution of the mass ratios, separations and circumbinary disc 
properties of such binaries as they accrete to their final masses.
Using this protobinary evolution code, we predict 
the properties of binary stars and how they depend
on the pre-collapse conditions in their progenitor molecular cloud
cores.  These predictions and their comparison with current observations
are discussed in detail in Section \ref{predictions}.

Briefly, we conclude that, if most binary stars form via the above model,
binary systems with smaller separations or greater total masses
should have mass ratios which are biased toward equal masses when 
compared to binaries with wider orbits or lower total masses.  Similarly,
the frequency of circumbinary discs should be greater for 
pre-main-sequence binaries with closer orbits or greater total masses.
These conclusions can be understood because: binaries which are closer or 
have a greater final mass should accrete more gas relative to their 
initial masses than wider or lower-mass binaries; 
the specific angular momentum of the infalling gas relative to that of
the binary is expected to increase as the accretion proceeds; 
and the accretion of gas with high specific angular momentum tends
to equalise the mass ratio and forms a circumbinary disc.
We also demonstrate that in a young binary which is accreting from 
an infalling gaseous envelope, the primary will generally have a
circumstellar disc which is more massive or similar in mass to that
of the secondary.  All of these
conclusions are in good agreement with the latest observations.

By making rough quantitative predictions of the 
properties of binary stars, we find that the observed properties 
of binary stars are most easily
reproduced if the pre-collapse molecular cloud cores from which binaries
form have radial density
profiles between uniform and $1/r$ (e.g.~Gaussian) with near 
uniform rotation.  This is in excellent agreement with the observed properties
of pre-stellar cores (Ward-Thompson et al.~1994;
Andr\'e, Ward-Thompson, Motte 1996; Ward-Thompson, Motte, \& Andr\'e 1999).  
Conversely, the observed properties of binaries cannot be 
reproduced if the cores are in solid-body rotation and 
have initial density profiles which are strongly centrally condensed
(between $1/r$ and $1/r^2$), and the singular isothermal sphere 
($\rho \propto 1/r^2$) cannot
fit the observations even with strong differential rotation.

\section*{Acknowledgments}

I am grateful to Ian Bonnell, Cathie Clarke and Jim Pringle for 
many helpful discussions and
their critical reading of the manuscript.

\end{document}